\documentclass[prd,twocolumn,superscriptaddress,nofootinbib,notitlepage]{revtex4-1} 

\usepackage[utf8]{inputenc}
\usepackage{hyperref}
\usepackage{amsfonts}
\usepackage{enumitem}
\hypersetup{
    colorlinks=true,
    linkcolor=blue,
    filecolor=magenta,      
    urlcolor=blue,
    pdfpagemode=FullScreen,
    citecolor=blue
}
\usepackage{verbatim}
\usepackage{upgreek}
\usepackage[normalem]{ulem}
\usepackage{graphicx}
\usepackage[dvipsnames]{xcolor}
\usepackage{soul}
\usepackage{ulem, color, framed}
\usepackage{amsmath}
\usepackage{mathrsfs}
\usepackage{subfigure}
\usepackage{braket}
\usepackage{cleveref}
\crefname{section}{Sec.}{Secs.}
\crefname{figure}{Fig.}{Figs.}
\crefname{appendix}{Appendix}{Appendices}
\crefname{equation}{Eq.}{Eqs.}
\crefname{table}{Table}{Tables}
\Crefname{section}{Section}{Sections}
\Crefname{figure}{Figure}{Figures}
\Crefname{appendix}{Appendix}{Appendices}
\Crefname{equation}{Equation}{Equations}
\Crefname{table}{Table}{Tables}
\usepackage{mathtools}
\usepackage{amssymb}
\usepackage{lineno}
\usepackage{float}
\usepackage{slashed}
\usepackage{longtable}
\usepackage{xfrac,bigints}
\usepackage{bm}
\usepackage{xspace}
\usepackage{color}
\usepackage{overpic}
\usepackage{pict2e}
\usepackage{multirow}
\usepackage[dvipsnames]{xcolor}

\newenvironment{sloppypar*}{\sloppy\ignorespaces}{\par}
\allowdisplaybreaks 

\newcommand {\sket} [1] {| #1 \rangle}

\newcommand{\pd}{{\vphantom{\dagger}}}

\newcommand{\GeV}{{{\,}\textrm{GeV}}}

\newcommand*\diff{\mathop{}\!\mathrm{d}}

\newcommand{\modes}{J}
\newcommand{\modesq}{\modes_q}
\newcommand{\modesa}{\modes_{\overline{q}}}
\newcommand{\modesg}{\modes_g}

\newcommand{\NF}{N_F}
\newcommand{\NB}{N_B}
\newcommand{\ind}{{\beta}} 
\newcommand{\occf}{{\omega}}
\newcommand{\occb}{{\widetilde{\omega}}}
\newcommand{\occmax}{\occb^{\mathrm{max}}}

\newcommand{\expconfig}[1]{{\langle #1 \rangle_{\text{event}}}}

\newcommand{\pops}{\vec{P}_\perp^2}

\newcommand{\floor}[1]{\left\lfloor #1 \right\rfloor}
\newcommand{\ceil}[1]{\left\lceil #1 \right\rceil}
\DeclareMathAlphabet{\mathpzc}{OT1}{pzc}{m}{it}

\begin{document}

\title{Efficient Quantum Simulation of QCD Jets on the Light Front}
\author{Wenyang Qian}
\email[]{qian.wenyang@usc.es}
\affiliation{
Instituto Galego de Fisica de Altas Enerxias (IGFAE), Universidade de Santiago de Compostela, E-15782 Galicia, Spain
}
\author{Meijian Li}
\email[]{meijian.li@usc.es}
\affiliation{
Instituto Galego de Fisica de Altas Enerxias (IGFAE), Universidade de Santiago de Compostela, E-15782 Galicia, Spain
}
\author{Carlos A. Salgado}
\email[]{carlos.salgado@usc.es}
\affiliation{
Instituto Galego de Fisica de Altas Enerxias (IGFAE), Universidade de Santiago de Compostela, E-15782 Galicia, Spain
}
\author{Michael Kreshchuk}
\email[]{michael@phasecraft.io}
\affiliation{
Physics Division, Lawrence Berkeley National Laboratory, Berkeley, California 94720, USA
}
\affiliation{
Phasecraft Inc, Washington, D.C. 20001, USA
}

\begin{abstract}
  Quark and gluon jets provide one of the best ways to probe the matter produced in ultrarelativistic high-energy collisions, from cold nuclear matter to hot quark-gluon plasma. 
  In this work, we propose a unified framework for efficient quantum simulation of many-body dynamics using the (3+1)-dimensional QCD Hamiltonian on the light front, particularly suited for studying the scattering of quark and gluon jets on nuclear matter in heavy-ion collisions.
  We describe scalable methods for mapping physical degrees of freedom onto qubits and for simulating in-medium jet evolution. 
  We then validate our framework by implementing an algorithm that directly maps second-quantized Fock states onto qubits and uses Trotterized simulation for simulating time dynamics. 
  Using a classical emulator, we investigate the evolution of quark and gluon jets with up to three particles in Fock states, extending prior studies. These calculations enable the study of key observables, including jet momentum broadening, particle production, and parton distribution functions.

\end{abstract}

\maketitle

\tableofcontents

\section{Introduction\label{sec:intro}}

Quantum simulation of high-energy physics (HEP) was one of the original motivations for the development of quantum computers~\cite{feynman1982simulating} and remains a source of ongoing proposals for achieving quantum advantage~\cite{Bauer:2022hpo, DiMeglio:2023nsa, Banuls:2019bmf}.
A key focus in HEP is the simulation of scattering processes~\cite{Jordan:2011ci,Jordan:2012xnu, Jordan:2014tma, Jordan:2017lea}, which are crucial for understanding particle interactions. Many directions have been extensively explored, ranging from simulation of gauge field theories~\cite{Hebenstreit:2013baa, Kasper:2015cca, Pichler:2015yqa,Martinez:2016yna,Klco:2018kyo,  Surace:2019dtp,Chakraborty:2020uhf,Magnifico:2020bqt,Ciavarella:2021lel,Nguyen:2021hyk, Bauer:2022hpo, Gonzalez-Cuadra:2022hxt,Farrell:2022wyt, Desaules:2022kse,Irmejs:2022gwv,Kreshchuk:2023btr, 
Belyansky:2023rgh, Florio:2023dke,Charles:2023zbl,Cataldi:2023xki,Calajo:2024qrc,Barata:2023jgd,Dempsey:2024alw,Barata:2024apg, Farrell:2024fit,Dempsey:2023gib,Ciavarella:2024cyt,Ciavarella:2025bsg,
Su:2024uuc, Halimeh:2023wrx,
Bennewitz:2024ixi,Davoudi:2024wyv,Zemlevskiy:2024vxt,Barata:2025hgx}, thermal systems~\cite{Buyens:2016ecr,Czajka:2021yll, Qian:2024xnr, Xie:2022jgj, Ikeda:2024rzv}, and thermalization of non-equilibrium systems~\cite{Lamm:2018siq, DeJong:2020riy,Mueller:2021gxd,Mueller:2024mmk} to extracting hadronic spectrum~\cite{Banuls:2013jaa,Choi:2020pdg,Ferguson:2020qyf,Kreshchuk:2020aiq,Atas:2021ext,Qian:2021jxp,Clemente:2022cka, Gallimore:2022hai,Rigobello:2023ype,Du:2024zvr}, parton distributions~\cite{Mueller:2019qqj,Lamm:2019uyc, Mueller:2019qqj, Echevarria:2020wct, Kreshchuk:2020dla, LiTianyin:2021kcs, LiTianyin:2024nod,Banuls:2024oxa,Barata:2024bzk,Grieninger:2024axp,Grieninger:2024cdl} and simulating parton showers~\cite{Bauer:2019qxa,Bauer:2021gup, Bepari:2021kwv, Gustafson:2022dsq}, among others. 
Experimentally, various efforts have also been made from tackling event reconstruction~\cite{Zlokapa:2019tkn,Magano:2021jzd} and jet clustering~\cite{Wei:2019rqy, Delgado:2022snu,deLejarza:2022vhe,deLejarza:2022bwc} to enhancing classical machine learning with quantum data encoding~\cite{Alvi:2022fkk, Ngairangbam:2021yma, Gianelle:2022unu,Delgado:2022tpc}.
Although most studies on quantum computing in high-energy physics (HEP) are limited by their simulation platforms---such as noisy intermediate-scale quantum (NISQ) devices~\cite{Preskill:2018jim}, quantum annealers~\cite{Yarkoni:2021zvu}, and classical simulators~\cite{Bergholm:2018cyq, cirq2024, qiskit2024, Bridgeman:2016dhh}---the proposition and formulation of problems using quantum computing represent a crucial step toward understanding the benefits and leveraging the strengths of future large-scale quantum computing applications.

Quark and gluon jets provide a powerful means of probing the matter and fields produced in ultrarelativistic high-energy collisions, from cold nuclear matter to the hot quark-gluon plasma~\cite{Busza:2018rrf}.
Jets are inherently quantum systems, and their interactions with a medium are highly complex, particularly their real-time dynamics.
Useful tools for studying such processes have been developed within the framework of light-front quantization.
The Discretized Light-Cone Quantization (DLCQ) approach~\cite{Pauli:1985pv, Pauli:1985ps, Eller:1986nt, Harindranath:1987db, Hornbostel:1988fb} has been developed as a powerful computational tool for investigating the properties of relativistic bound states.
It was later generalized to the Basis Light-Front Quantization (BLFQ) approach~\cite{1stBLFQ}, which employs a basis function representation to improve computational efficiency.
BLFQ has been successfully applied to address both QED systems~\cite{Honkanen:2010rc, Zhao:2014xaa, Wiecki:2014ola, Chakrabarti:2014cwa, Hu:2020arv, Nair:2022evk, Nair:2023lir} and QCD bound states~\cite{Jia:2018ary, Lan:2019vui, Lan:2019rba, Adhikari:2021jrh, Lan:2021wok, Mondal:2021czk, Li:2015zda, Li:2017mlw, Li:2018uif, Lan:2019img, Tang:2018myz, Tang:2019gvn, Mondal:2019jdg, Qian:2020utg, Xu:2021wwj, Liu:2022fvl, Hu:2022ctr, Peng:2022lte, Kaur:2023lun, Zhu:2023nhl, Zhang:2023xfe, Liu:2024umn}, with recent studies extending beyond the valence Fock sector to incorporate gluon dynamics~\cite{Kaur:2024iwn, Xu:2023nqv, Lin:2023ezw, Zhu:2023lst, Yu:2024mxo}.
The time-dependent Basis Light-Front Quantization (tBLFQ) approach\footnote{Its non-relativistic counterpart---the time-dependent Basis Function (tBF) approach--has been developed to address nuclear structure and scattering~\cite{Du:2018tce, Yin:2022zii}.} extends the BLFQ framework to tackle time-dependent problems in the presence of an external background field, first applied in QED~\cite{Zhao:2013cma, Chen:2017uuq, Hu:2019hjx, Lei:2022nsk}, and later in QCD for jet evolution~\cite{Li:2020uhl, Li:2021zaw, Li:2023jeh}.
However, classical calculations require resources that grow exponentially with momentum (or basis) cutoffs and the number of particles, limiting current developments to the leading Fock sectors. In contrast, quantum computing, by leveraging quantum parallelism, has the potential to reduce resource scaling to a polynomial level in terms of qubits or quantum operations relative to the problem size, presenting a promising approach for more efficient Hamiltonian simulation.

Recent work~\cite{Barata:2021yri, Barata:2022wim, Barata:2023clv, Wu:2024adk} has explored a non-perturbative quantum simulation framework for studying in-medium evolution of a quark jet with up to one dynamical gluon.
A compact \emph{basis state encoding} scheme efficiently reduces qubit usage, scaling logarithmically with the number of classical basis states. However, extending this approach to two or more gluons becomes challenging, as the Hamiltonian complexity grows exponentially with the number of basis states.
To address this challenge, we propose a unified quantum simulation strategy for quark and gluon jets as multi-particle QCD states.
The foundation of our method remains the non-perturbative light-front Hamiltonian formalism~\cite{Brodsky:1997de, 1stBLFQ, Li:2020uhl, Li:2021zaw, Li:2023jeh}, as used in previous work~\cite{Barata:2021yri, Barata:2022wim, Barata:2023clv}. 
In this formalism, the jet state is treated as a fully quantum state, and the simulation of the jet evolution is performed on the amplitude level.
We carry out the real-time evolution of the jet state by decomposing the time-evolution operator into time increments and letting the operation of each timestep act sequentially to the initial state. Such Trotterized calculation is non-perturbative and enables us to relax approximations usually made in other approaches, such as the  eikonal approximation (e.g., Ref.~\cite{Dumitru:2002qt}) and the multiple soft scattering approximation~\cite{Baier:1996kr, Baier:1996sk, Zakharov:1997uu}. 
In addition, performing real-time simulation enables intermediate state measurements, providing direct access to real-time observables.

Our current quantum simulation strategy employs a \emph{direct encoding} scheme: the qubit degrees of freedom store occupancies of fermionic~\cite{Jordan:1928wi, Bravyi:2000vfj} and bosonic modes~\cite{aspuru2005simulated}, enabling a straightforward mapping of the second-quantized light-front Hamiltonian to qubit operators.
Consequently, the number of quantum gate operations and qubit locality scale polynomially with system size, leading to an efficient and scalable algorithm.
This approach allows us to simulate jet evolution with multiple-gluon components from first principles, surpassing the capabilities of previous basis state encoding methods. 
To demonstrate the capability and feasibility of our algorithm, we simulate the in-medium evolution of both quark and gluon jets on a quantum circuit simulator.
Although our encoding scheme is generic and not constrained to a fixed number of particle sectors, we illustrate it using resources at currently available scales (i.e., 36-128 qubits) by considering three leading Fock sectors, $\ket{q}+\ket{qg}+\ket{qgg}$ and $\ket{g}+\ket{gg}+\ket{ggg}$ for the quark and gluon jets, respectively. 
This calculation is analytically challenging and extends beyond the scope of previous classical simulations.
We observe the dynamic interplay between the jet, the medium, and self-interactions, examining effects such as medium-induced modifications on the transverse momentum broadening, average gluon number production, and jet parton probability distribution.

The paper is organized as follows. In~\cref{sec:general_framework},
we review and extend the general formulation of in-medium jet evolution in the light-front Hamiltonian formalism, and specify the phase space and the problem Hamiltonian for the simulation. In~\cref{sec:qsim}, we introduce and discuss various aspects of the quantum simulation algorithm, especially on the encoding strategy. In~\cref{sec:results}, we present the numerical simulation results of the real-time evolution of the quark and gluon jets using {\texttt Qiskit}~\cite{qiskit2024} quantum simulator. In Section.~\ref{sec:summary}, we summarize and discuss future avenues of research.

\section{QCD Jets in the Light-Front Hamiltonian formalism \label{sec:general_framework}}

In this section, we review and extend the theoretical formulation of a high-energy parton evolving through an external background field representing the medium, using a non-perturbative light-front (LF) Hamiltonian formalism.
Our presentation closely follows the established Discretized Light-Cone Quantization (DLCQ)~\cite{Brodsky:1997de} and time-dependent Basis Light-Front Quantization (tBLFQ) approaches~\cite{Li:2020uhl, Li:2021zaw, Li:2023jeh}.
Previous studies ~\cite{Li:2020uhl, Li:2021zaw, Li:2023jeh, Barata:2022wim, Barata:2023clv} simulated jet quark evolution in the Fock space up to $\ket{q}+\ket{qg}$. Since jets are inherently many-particle states produced through branching processes of an initial particle, we aim to extend the framework to describe both quark and gluon jets as many-particle states using a scalable algorithm.

\begin{figure}[t]
  \centering
  \includegraphics[width=0.38\textwidth]{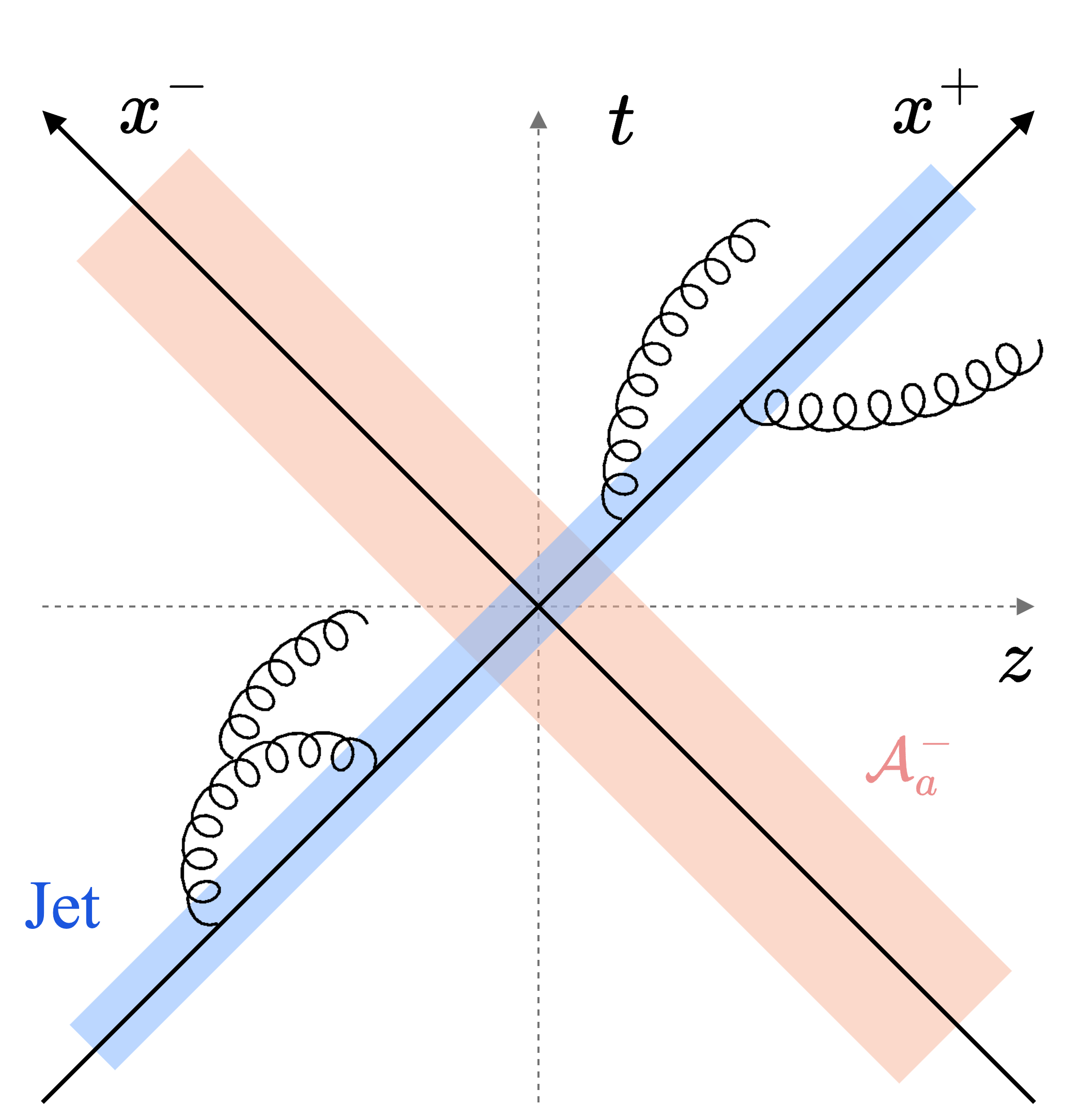}
  \caption{Schematic of a jet probe scattering off a nucleus in a spacetime diagram.
  }
 \label{fig:dis_zt}
\end{figure}

We consider a high-energy jet, produced by either a quark or gluon, moving in the positive $z$-direction. The jet scatters off a background field, which moves in the negative $z$-direction and represents cold nuclear matter from a high-energy nucleus or the dense medium created in heavy-ion collisions, as illustrated in~\cref{fig:dis_zt}.
The jet has momentum $p^\mu$ and $p^+\gg p^-, p_\perp$ whereas the nucleus/medium has momentum $P^\mu$ and $P^-\gg P^+, P_\perp$ (see definitions of the light-front variables in~\cref{app:convention}).
We treat the jet as a fully quantum state governed by QCD dynamics, with the nucleus/medium described as a background gluon field.
The quark interacts with the background field for a finite duration of $0\le x^+\le L_\eta$. 

\subsection{The full Hamiltonian: QCD + Background field \label{ssec:general}}

The Lagrangian for the process under consideration is the QCD Lagrangian with an external background field $\mathcal{A}^\mu$,
\begin{align}\label{eq:Lagrangian}
 \mathcal{L}=-\frac{1}{4}{F^{\mu\nu}}_a F^a_{\mu\nu}+\bar{\Psi}(i\gamma^\mu  D_\mu -  m_q)\Psi\;,
\end{align}
where $F^{\mu\nu}_a\equiv\partial^\mu C^\nu_a-\partial^\nu C^\mu_a-g f^{abc}C^\mu_b C^\nu_c$ is the field strength tensor, $D^\mu\equiv \partial_\mu +ig C^\mu$ the covariant derivative, and $ C^\mu= A^\mu + \mathcal{A}^\mu$ is the sum of the quantum gauge field $ A^\mu$ and the background gluon field $\mathcal{A}^\mu$. The light-front Hamiltonian $P^-$ can be derived from the Lagrangian in~\cref{eq:Lagrangian} through the standard Legendre transformation~\cite{Brodsky:1997de, Li:2021zaw} and acquires the form of
\begin{equation}
    \label{eq:Hamiltonian}
    P^-(x^+) =P^-_\textrm{QCD}+ V_{\mathcal{A}}(x^+) \;,
\end{equation}
Here, $P^-_{QCD}$ is the vacuum QCD Hamiltonian, i.e., the standard QCD Hamiltonian without the background field and $V_{\mathcal{A}}(x^+)$ contains the interactions with the background field $ \mathcal A$. These are written as:
\begin{align}
\label{eq:PQCD}
     P^-_{QCD} =P^-_{\text{KE}} + V_{qg}+V_{ggg}+V_{gggg}+W_{g} + W_{f}\;,
\end{align}
where $P^-_{\text{KE}}$ denotes the kinetic energy term, $V_{qg}$, $V_{ggg}$, and $V_{gggg}$ represent the quark-gluon and multi-gluon interaction vertices, and $W_g$ and $W_f$ are the gluon and fermion instantaneous terms, respectively.
The interaction with the background field $\mathcal{A}$ is given by:
\begin{equation}
    V_{\mathcal{A}}(x^+)=V_{q\mathcal{A}}(x^+) + V_{g\mathcal{A}}(x^+)\;,
\end{equation}
where $V_{q\mathcal{A}}(x^+)$ and $V_{g\mathcal{A}}(x^+)$ represent the quark-field and gluon-field interactions with the background field $\mathcal{A}$, respectively.

The quantization of the field is performed at equal light-front time \( x^+ \) in a discrete 3-dimensional momentum representation.
Upon substituting the mode expansions of the fields, the Hamiltonian acquires the form of a polynomial in bosonic and fermionic creation and annihilation operators [see~\cref{eq:PKE,eq:VgA}] acting on the Fock states spanning the Hilbert space of the model.
Here, we summarize the discretization scheme and introduce the relevant parameter notations, while leaving the details to \cref{app:modes}. 
The dimensionless discrete longitudinal momentum quanta of the single-particle mode is \( k^+ = 1/2, 3/2, \ldots \) (\(1, 2, \ldots\)) for fermions (bosons). 
The maximal value is restricted by the \emph{harmonic resolution} \( K = P^+ L/(2\pi) \), where \( P^+ \) is the total longitudinal momentum of the system and \( 2L \) the extent of \( x^- \).
The unit of the physical \( p^+ \) is therefore \( d_+ = 2\pi/L \), so that \( p^+ = k^+ d_+ \).
The transverse momentum is discretized on a square lattice, with dimensionless quanta \( k^i = -N_\perp, -N_\perp+1, \ldots, N_\perp -1 \) (\(i=x,y\)). 
The transverse position space is also discretized with periodic boundary conditions imposed, and the extent of the space is \( 2 L_\perp \).
Consequently, the resolution in the transverse position space is \( a_\perp \equiv L_\perp/N_\perp \), and in the transverse momentum space is \( d_p = \pi/L_\perp \), such that the physical transverse momentum can be expressed as \( p^j = k^j d_p \). The ultraviolet (UV) cutoff in the transverse momentum space is \( \lambda_{UV} = \pi/a_\perp \).

\subsubsection{The Hilbert space of jets}

In general, the QCD Fock states constituting the Hilbert space of a many-particle system have the form
\begin{equation}
\label{eq:fock_general}
\sket{
q_1,\,q_2,\ldots;
\overline{q}_1,\,\overline{q}_2,\ldots;
g_1,\,g_2,\ldots
}\;,
\end{equation}
where $q_j$, $\overline{q}_j$, and $g_j$ represent the quark, antiquark, and gluon single-particle states, with $j$ denoting the index of different \emph{modes}.
In the chosen discrete momentum basis states, each mode is a single-particle state labeled by five quantum numbers, collectively represented by the \emph{mode index}:
\begin{align}\label{eq:basis_mode}
      \ind=\{ k^+, k^x, k^y, \lambda,c(a) \}\;,
  \end{align}
where $\lambda$ is the light-front helicity, and $c=1,2,\ldots, N_c$ ($a=1,2,\ldots N_c^2-1 $) is the color index of quark/antiquark (gluon); see Ref.~\cite{Brodsky:1997de} and~\cref{app:Pmin}. 

The formation of a jet originates from an initially highly energetic quark or gluon undergoing a branching process dominated by gluon emission and absorption, before eventually hadronizing. Our focus is on simulating and studying the interplay between medium interactions and the branching process, where multiple gluons are produced. 
As a result, we consider the Fock space of the quark and gluon jets expressed as
\begin{subequations}
\label{eq:jet_states}
\begin{alignat}{9}
\label{eq:jet_state_quark}
&
\ket{q}_{\mathrm{jet}} && = \psi_q\ket{q}+\psi_{qg}\ket{qg}+\psi_{qgg}\ket{qgg}+\cdots
\\
\label{eq:jet_state_gluon}
&\ket{g}_{\mathrm{jet}}&&=\psi_g\ket{g}+\psi_{gg}\ket{gg}+\psi_{ggg}\ket{ggg}+\cdots
\end{alignat}
\end{subequations}
With this approximation, we neglect contributions that involve antiquarks such as the antenna process of $ g\to q+ \bar q$, which are less dominant for the current study but may become important in the later stage of fragmentation.

In practice, performing numerical calculations requires truncating the infinite-dimensional Hilbert space to a finite-dimensional one. 
To achieve this, we impose cutoffs on the single-particle quantum numbers, specifically, the number of discretized momentum modes, as well as the number of gluons occupying the same mode, as will be discussed in more detail in~\cref{sec:qsim}.
The chosen truncation treatment is similar to that implemented in DLCQ~\cite{Brodsky:1997de, Brodsky:2014yha, Harindranath:1987db} and BLFQ~\cite{1stBLFQ, Vary_2018}, albeit with slight variations.
The dimension of the resulting Hilbert space grows exponentially with the maximum number of particles in a Fock state, thereby enabling the advantages of quantum simulation.

In the considered Fock space of the jets, as described in ~\cref{eq:jet_states}, the relevant contributing terms of the Hamiltonian in ~\cref{eq:PQCD} can be written as follows:
\begin{subequations}
\label{eq:Pm}
\begin{alignat}{9}
\label{eq:Pmn_qjet}
&\begin{alignedat}{9}
P_{q ~\mathrm{jet}}^-(x^+)=&
    K_q + K_g
    + V_{qg}
    + V_{\mathcal{A}}(x^+)
     +V_{ggg}\\
     &
     +V_{gggg}
      +\sum_{i=1}^5 W_{g,i}
     +\sum_{j=1}^3 W_{q,j}
    \;,
\end{alignedat}
\\
\label{eq:Pmn_gjet}
&\begin{alignedat}{9}
P_{g ~\mathrm{jet}}^-(x^+)=&
    K_q + K_g
    + V_{qg}
    + V_{\mathcal{A}}(x^+)
     +V_{ggg}\\
     &
     +V_{gggg}
     +\sum_{i=1}^3 W_{g,i}
    \;.    
\end{alignedat}
\end{alignat}
\end{subequations}

We list the terms of the above Hamiltonian operators in their second-quantized form in the discretized momentum basis in~\cref{tab:second_quantized,tab:second_quantized_cont}, where $b_\beta^\dagger$ ($b_\beta$) is the fermionic creation (annihilation) operator and $a^\dagger_\beta$ ($a_\beta$) is the bosonic creation (annhilation) operator for a given mode $\beta$ as defined in \cref{eq:basis_mode}. See~\cref{app:Pmin} for further details. 
These tables will also serve as a convenient reference for implementing quantum simulations at the operator level.

\renewcommand{\arraystretch}{2}
\begin{table*}
\setlength\tabcolsep{8pt}
 \centering
 \caption{Groups of terms entering the Hamiltonian operators in~\cref{eq:Pm} in the second-quantized form for the QCD jet evolution in a background field $\mathcal{A}(x^+)$.
 }
 \label{tab:second_quantized} 
 \begin{tabular}{ |c|c|  c|  c|  c| } 
 \hline
\multicolumn{2}{|c|}{ Hamiltonian }
 & Diagram
 & Coefficient
 & Operator\\ 
 \hline
\multirow{2}{*}{  $P^-_{\text{KE}}$ }
  &$K_q$
    &
  ---
 & 
   $\big(p_\perp^2 + m_q^2\big)/p^+$
  & 
    $b_\beta^\dagger b^\pd_{\beta}$
 \\
 \cline{2-5}
 &$K_g$
 & 
 \begin{tabular}{c}
   ---
 \end{tabular}
 & 
 \begin{tabular}{c}
   $p_\perp^2 /p^+$ 
 \end{tabular}
  & 
 \begin{tabular}{c}
    $a_\beta^\dagger a\pd_\beta$
 \end{tabular}
 \\
 \hline
\multirow{2}{*}{  $V_{\mathcal A}$ }
&
$V_{q\mathcal A}$
 &  \begin{minipage}[c]{.15\textwidth}
        \centering
       ~~ \\
      ~~ \\
     \begin{overpic}[width=\textwidth]{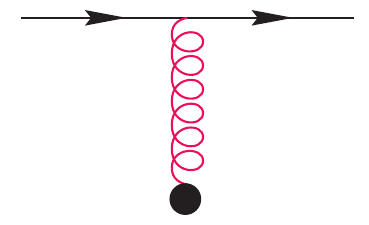 }
    \put (-10,50) {$1$}
    \put (100,50) {$2$}
    \end{overpic}
    ~~ \\
    \end{minipage}
 & 
 \begin{tabular}{c}
 $\dfrac{2g}{\Omega_\perp }
    \delta_{k_2^+,k_1^+}
    \delta_{\lambda_1,\lambda_2}
     T_{c_2,c_1}^a\tilde{\mathcal{A}}^a_+(\vec p_{\perp,2}- \vec p_{\perp,1}, x^+)$
 \end{tabular}
  & 
 \begin{tabular}{c}
    $b_{\beta_2}^\dagger b\pd_{\beta_1}$
 \end{tabular}
 \\
 \cline{2-5}
 &
$ V_{g\mathcal A}$
 &    \begin{minipage}[c]{.15\textwidth}
        \centering
       ~~ \\
      ~~ \\
     \begin{overpic}[width=\textwidth]{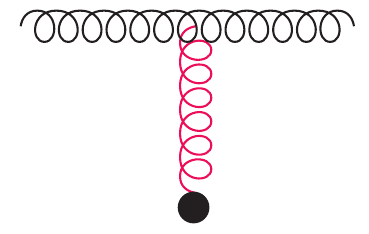 }
    \put (-10,50) {$1$}
    \put (100,50) {$2$}
    \end{overpic}
    ~~\\
    \end{minipage}
 & 
 \begin{tabular}{c}
  $-\dfrac{ i 2 g }{\Omega_\perp}
\delta_{k_1^+,k_2^+}
\delta_{\lambda_1,\lambda_2}
f^{a a_2 a_1}
\tilde{\mathcal{A}}^a_+(\vec p_{\perp,2}- \vec p_{\perp,1}, x^+)$
 \end{tabular}
  & 
 \begin{tabular}{c}
    $a^\dagger_{\beta_2} a\pd_{\beta_1}$
 \end{tabular}
 \\
 \hline
\multicolumn{2}{|c|}{ $V_{qg}$}
 & 
\begin{minipage}[c]{.15\textwidth}
        \centering
       ~~ \\
      ~~ \\
     \begin{overpic}[width=\textwidth]{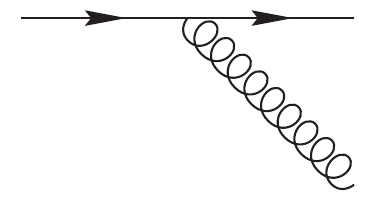 }
    \put (-10,45) {$1$}
    \put (100,45) {$2$}
    \put (100,0) {$3$}
    \end{overpic}
    ~~ \\
    \end{minipage}
 & 
 $
 \dfrac{g }{\sqrt{ p_1^+ p_2^+ p_3^+ \Omega}}
   T_{c_2,c_1}^{a_3}
  \delta^{(3)}_{p_2 - p_1 +p_3}
   \Delta_1^{2,3}
 $
  & 
    $ b^\dagger_{\beta_2} b^\pd_{\beta_1}a^\dagger_{\beta_3}  $
 \\
\hline
 \multicolumn{2}{|c|}{$V_{ggg}$}
 & 
\begin{minipage}[c]{.15\textwidth}
        \centering
       ~~ \\
      ~~ \\
     \begin{overpic}[width=\textwidth]{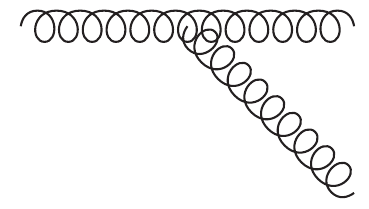 }
    \put (-10,45) {$1$}
    \put (100,45) {$2$}
    \put (100,0) {$3$}
    \end{overpic}
    ~~ \\
    \end{minipage}
 & 
 \begin{tabular}{c}
 $ 
    \dfrac{-ig }{\sqrt{p_1^+p_2^+ p_3^+ \Omega}}
    \delta^{(3)}_{p_1-p_2-p_3}
    f^{a_1 a_2 a_3}
    \Sigma_1^{2,3}$
 \end{tabular}
  & 
 \begin{tabular}{c}
    $a^\dagger_{\beta_2} a^\dagger_{\beta_3} a\pd_{\beta_1}$
 \end{tabular}
 \\
 \hline
\multirow{2}{*}{ 
 \begin{tabular}{c}
 \\
 \\
$ V_{gggg}$
 \end{tabular}
}
& $ V_{gggg,1}$
 & 
\begin{minipage}[c]{.15\textwidth}
        \centering
       ~~ \\
      ~~ \\
     \begin{overpic}[width=\textwidth]{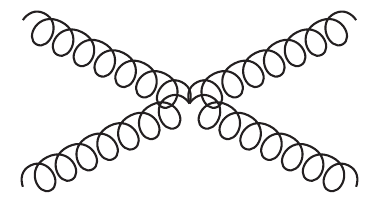 }
    \put (-10,45) {$1$}
    \put (-10,0) {$2$}
    \put (100,45) {$3$}
    \put (100,0) {$4$}
    \end{overpic}
    ~~ \\
    \end{minipage}
 & 
$ \begin{array}{r@{}l@{}}
& \\
  \dfrac{g^2}{2}  &
    \dfrac{1}{\sqrt{ p_1^+ p_2^+p_3^+ p_4^+} \Omega}
    \delta^3_{p_1+p_2-p_3-p_4}\\
   ~& \bigg[
    f^{a a_1 a_2}  f^{a a_3 a_4} \delta_{\lambda_1, \lambda_3}
    \delta_{\lambda_2, \lambda_4}\\
   ~& +f^{a a_1 a_3}  f^{a a_2 a_4}  \delta_{\lambda_1, -\lambda_2}
    \delta_{\lambda_3, -\lambda_4}\\
   ~&
    +f^{a a_1 a_4}  f^{a a_3 a_2} \delta_{\lambda_1, \lambda_3}
    \delta_{\lambda_2, \lambda_4}
    \bigg] \\
    &
 \end{array}$
  & 
    $ a^\dagger_{\beta_4} a^\dagger_{\beta_3} a^\pd_{\beta_2} a^\pd_{\beta_1} $
 \\
 \cline{2-5}
  & $ V_{gggg,2}$
 &\begin{minipage}[c]{.15\textwidth}
        \centering
       ~~ \\
      ~~ \\
     \begin{overpic}[width=\textwidth]{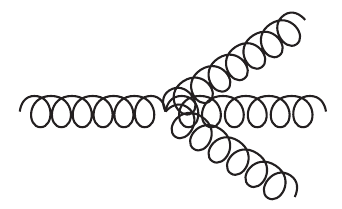 }
    \put (-10,25) {$1$}
    \put (100,50) {$2$}
    \put (100,25) {$3$}
    \put (100,0) {$4$}
    \end{overpic}
    ~~ \\
    \end{minipage} 
    &
$\begin{array}{r@{}l@{}}
g^2&
\dfrac{1}{\sqrt{ p_1^+ p_2^+p_3^+ p_4^+} \Omega}
\delta^3_{p_1-p_2-p_3-p_4}\\
&
f^{a a_1 a_2}  f^{a a_3 a_4} \delta_{\lambda_1, \lambda_3}
\delta_{\lambda_2, -\lambda_4}
\end{array}$
    &
$a_{\beta_4}^\dagger a_{\beta_3}^\dagger a_{\beta_2}^\dagger a^\pd_{\beta_1}$
    \\
 \hline
 \end{tabular}
\end{table*}

\begin{table*}[t]
\setlength\tabcolsep{8pt}
 \centering
 \caption{Continuation of Table~\ref{tab:second_quantized}, the instantaneous terms.  }
 \label{tab:second_quantized_cont} 
 \begin{tabular}{| c | c | c | c | c |} 
 \hline
 \multicolumn{2}{|c|}{ Hamiltonian }
 & Diagram
 & Coefficient
 & Operator\\ 
 \hline
\multirow{5}{*}{
 \begin{tabular}{c}
 \\
 ~~\\
 ~~\\
 ~~\\
  ~~\\
 ~~\\
   $W_{g}$
 \end{tabular}
} 
 & 
  $W_{g,1 }$
  &
\begin{minipage}[c]{.15\textwidth}
        \centering
       ~~ \\
      ~~ \\
     \begin{overpic}[width=\textwidth]{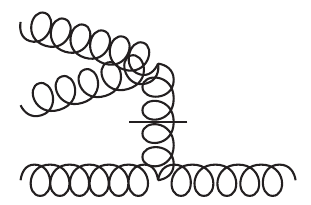 }
    \put (-10,50) {$1$}
    \put (-10,30) {$2$}
    \put (-10,0) {$3$}
    \put (100,0) {$4$}
    \end{overpic}
    ~~ \\
    \end{minipage}
 & 
$ \begin{array}{r@{}l@{}}
 - g^2 &
       \delta^3_{p_1+p_2+p_3-p_4}
       \dfrac{1}{\sqrt{ p_1^+ p_2^+p_3^+ p_4^+}\Omega}\\
     ~ & f^{a a_1 a_2} f^{a a_3 a_4}
        \dfrac{ p_1^+ (p_3^+ + p_4^+)}{{(p_3^+ -p_4^+ )}^2}
       \delta_{\lambda_1, -\lambda_2}
       \delta_{\lambda_3, \lambda_4}  
 \end{array}$
  & 
    $ a^\dagger_{\beta_4} a^\pd_{\beta_1} a^\pd_{\beta_2} a^\pd_{\beta_3} $
 \\
 \cline{2-5}
  & 
  $W_{g,2 }$
 &\begin{minipage}[c]{.15\textwidth}
        \centering
       ~~ \\
      ~~ \\
     \begin{overpic}[width=\textwidth]{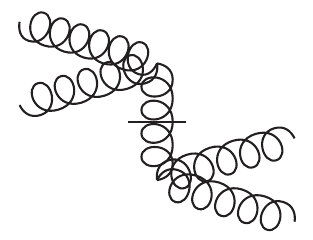 }
    \put (-10,65) {$1$}
    \put (-10,40) {$2$}
    \put (100,25) {$3$}
    \put (100,0) {$4$}
    \end{overpic}
    ~~ \\
    \end{minipage} 
    &
$\begin{array}{r@{}l@{}}
g^2&
        \delta^3_{p_1+p_2-p_3-p_4}
        \dfrac{1}{\Omega}
        \sqrt{ \dfrac{p_1^+ p_3^+ }{ p_2^+ p_4^+} }
        f^{a a_1 a_2} f^{a a_3 a_4}\\
      ~ &
     \dfrac{1}{{(p_3^+ +p_4^+ )}^2}
     \delta_{\lambda_1, -\lambda_2}
     \delta_{\lambda_3, -\lambda_4}
\end{array}$
    &
$a_{\beta_3}^\dagger a_{\beta_4}^\dagger a^\pd_{\beta_1}  a^\pd_{\beta_2} $
    \\
     \cline{2-5}
        & 
  $W_{g,3 }$
 &\begin{minipage}[c]{.15\textwidth}
        \centering
       ~~ \\
      ~~ \\
     \begin{overpic}[width=\textwidth]{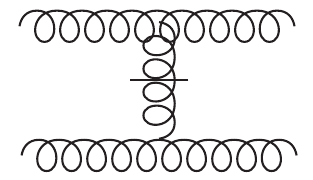 }
    \put (-10,0) {$2$}
    \put (-10,50) {$1$}
    \put (100,50) {$3$}
    \put (100,0) {$4$}
    \end{overpic}
    ~~ \\
    \end{minipage} 
    &
    $\begin{array}{r@{}l@{}}
-\dfrac{g^2}{2}&
           \delta^3_{p_1+p_2-p_3-p_4}
           \dfrac{1}{\sqrt{ p_1^+ p_2^+p_3^+ p_4^+} \Omega}
           f^{a a_1 a_3} f^{a a_2 a_4}
           \\
           ~&
          \dfrac{(p_1^+ + p_3^+)( p_2^+ + p_4^+)}{{(p_2^+ -p_4^+ )}^2}
          \delta_{\lambda_1, \lambda_3}
          \delta_{\lambda_2, \lambda_4}
\end{array}$
    &$a_{\beta_3}^\dagger a_{\beta_4}^\dagger a^\pd_{\beta_1}  a^\pd_{\beta_2} $
    \\
     \cline{2-5}
        & 
  $W_{g,4 }$
 &\begin{minipage}[c]{.15\textwidth}
        \centering
       ~~ \\
      ~~ \\
     \begin{overpic}[width=\textwidth]{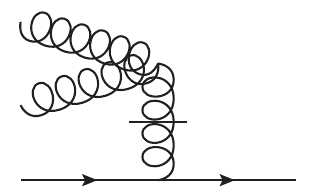 }
    \put (-10,0) {$3$}
    \put (-10,50) {$1$}
    \put (-10,25) {$2$}
    \put (100,0) {$4$}
    \end{overpic}
    ~~ \\
    \end{minipage} 
    &
        $\begin{array}{r@{}l@{}}
    2 i g^2&
\delta^3_{p_1+p_2+p_3-p_4}
\dfrac{1}{\Omega}
\sqrt{ \dfrac{p_1^+} {p_2^+}}
f^{a a_1 a_2} T^a_{c_4, c_3}\\
~&
\dfrac{1}{{(p_3^+ -p_4^+)}^2}
\delta_{\lambda_1,-\lambda_2}
\delta_{\lambda_3,\lambda_4}
\end{array}$
    &$b_{\beta_4}^\dagger b^\pd_{\beta_3} a^\pd_{\beta_1} a^\pd_{\beta_2}   $
    \\
     \cline{2-5}
        & 
  $W_{g,5 }$
 &\begin{minipage}[c]{.15\textwidth}
        \centering
       ~~ \\
      ~~ \\
     \begin{overpic}[width=\textwidth]{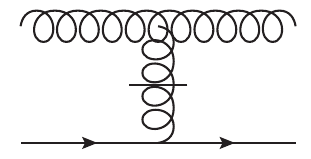 }
    \put (-10,0) {$2$}
    \put (-10,40) {$1$}
    \put (100,40) {$3$}
    \put (100,0) {$4$}
    \end{overpic}
    ~~ \\
    \end{minipage} 
    &
        $\begin{array}{r@{}l@{}}
   2 i g^2 &
    \delta^3_{p_1+p_2-p_3-p_4}
    \dfrac{1}{\sqrt{ p_1^+ p_3^+ }\Omega}
    f^{a a_1 a_3} T^a_{c_4, c_2}\\
  ~ &
 \dfrac{p_1^+ +p_3^+ }{{(p_4^+ - p_2^+)}^2}
   \delta_{\lambda_1,\lambda_3}
   \delta_{\lambda_2,\lambda_3}
\end{array}$
    &$ b_{\beta_4}^\dagger b^\pd_{\beta_2} a_{\beta_3}^\dagger   a^\pd_{\beta_1}   $
    \\
 \hline
      & 
  $W_{q,1 }$
 &\begin{minipage}[c]{.15\textwidth}
        \centering
       ~~ \\
      ~~ \\
     \begin{overpic}[width=\textwidth]{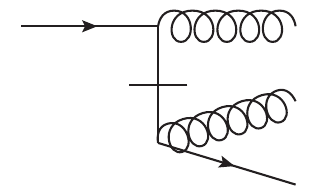 }
    \put (-10,50) {$1$}
    \put (100,50) {$4$}
    \put (100,25) {$3$}
    \put (100,0) {$2$}
    \end{overpic}
    ~~ \\
    \end{minipage} 
    &
    $ \begin{array}{r@{}l@{}}
     2 g^2&
      \delta^3_{p_2+p_3+p_4-p_1}
      \dfrac{1}{\sqrt{ p_3^+ p _4^+}\Omega}
      T^{a_3}_{c_2,c}  T^{a_4}_{c, c_1}\\
      ~&
      \dfrac{1}{p_1^+ - p_4^+}
      \delta_{\lambda_2,\lambda_3}
      \delta_{\lambda_4,-\lambda_1}
      \delta_{\lambda_2,\lambda_1}
       \end{array}$
    &$b^\dagger_{\beta_2} b^\pd_{\beta_1} a_{\beta_3}^\dagger a_{\beta_4}^\dagger   $
    \\
     \cline{2-5}
\multirow{3}{*}{
 \begin{tabular}{c}
   $W_f$\\
   ~\\
    ~\\
 \end{tabular}
}   & 
  $W_{q,2 }$
 & 
\begin{minipage}[c]{.15\textwidth}
        \centering
       ~~ \\
      ~~ \\
     \begin{overpic}[width=\textwidth]{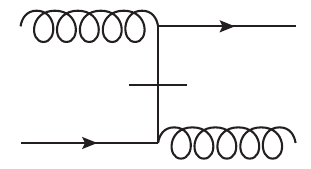 }
        \put (-10,50) {$2$}
        \put (-10,0) {$1$}
        \put (100,50) {$3$}
        \put (100,0) {$4$}
    \end{overpic}
    ~~ \\
    \end{minipage}
 & 
$ \begin{array}{r@{}l@{}}
          2g^2&
          \delta^3_{p_3-p_2+p_4-p_1}
          \dfrac{1}{\sqrt{  p_2^+ p_4^+ }\Omega}
          T^{a_2}_{c_3,c}  T^{a_4}_{c, c_1}\\
          ~&
          \dfrac{1}{p_1^+ - p_4^+}
          \delta_{\lambda_3,-\lambda_2}
         \delta_{\lambda_3,-\lambda_4}
         \delta_{\lambda_3,\lambda_1}
 \end{array}$
  & 
    $ b_{\beta_3}^\dagger  b^\pd_{\beta_1}a^\dagger_{\beta_4} a^\pd_{\beta_2}  $
 \\
 \cline{2-5}
        & 
  $W_{q,3 }$
 &\begin{minipage}[c]{.15\textwidth}
        \centering
       ~~ \\
      ~~ \\
     \begin{overpic}[width=\textwidth]{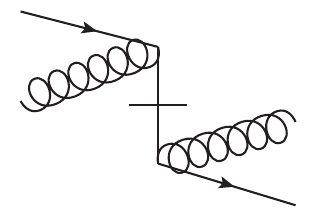 }
        \put (-10,60) {$1$}
        \put (-10,35) {$2$}
        \put (100,25) {$4$}
        \put (100,0) {$3$}
    \end{overpic}
    ~~ \\
    \end{minipage} 
    &
$ \begin{array}{r@{}l@{}}
          2g^2
&         \delta^3_{p_3+p_4-p_2-p_1}
         \dfrac{1}{\sqrt{  p_4^+ p_2^+ }\Omega}
        T^{a_4}_{c_3,c}  T^{a_2}_{c, c_1}\\
        ~&
        \dfrac{1}{p_2^+ + p_1^+}
        \delta_{\lambda_3,\lambda_4}
       \delta_{\lambda_3,\lambda_2}
       \delta_{\lambda_3,\lambda_1}
 \end{array}$
    &$b_{\beta_3}^\dagger b^\pd_{\beta_1} a_{\beta_4}^\dagger   a^\pd_{\beta_2} $
    \\
 \hline
 \end{tabular}
\end{table*}

\subsubsection{The background field}\label{sec:background_A}
The background field $\mathcal{A}^\mu$ accounts for the target nucleus or medium, and we describe it using the McLerran--Venugopalan (MV) model~\cite{McLerran:1993ni,McLerran:1993ka,McLerran:1998nk}, as in the preceding works~\cite{Li:2020uhl, Li:2021zaw, Li:2023jeh,Barata:2021yri,Barata:2022wim, Barata:2023clv}.
In the MV model, one assumes that the field has only one nonzero component, $\mathcal{A}^-$, and is independent of its light-front time $x^-$, as the field possesses a large $P^-$.
The field is formulated as a classical gluon field, and can be solved from the reduced Yang-Mills equation,
\begin{align}\label{eq:poisson}
 (m_g^2-\nabla^2_\perp )  \mathcal{A}^-_a(\vec{x}_\perp,x^+)=\rho_a(\vec{x}_\perp,x^+)\;,
\end{align}
where $m_g$ is the effective gluon mass of the field introduced to regularize the infrared (IR) divergence.
The source color charge density obeys a local correlation function
\begin{multline}\label{eq:MV_color_charge}
 \langle \rho_a(x^+,\vec x_\perp)\rho_b({x'}^{+},\vec x'_\perp) \rangle_{\text{event}} =\\
 g^2 \mu^2\delta_{ab}\,\delta^{(2)}(\vec x_\perp-\vec x'_\perp)\,\delta(x^+-{x'}^+)\;,
\end{multline}
where $\langle \cdot \rangle_{\text{event}}$ denotes the configuration average, and $\mu$ the medium strength. 
For a given $\mu$, the equation holds after averaging over multiple events, with each event having a different configuration of $\rho$ in the $\{x^+,\vec x_\perp\}$ space. Physically, this corresponds to the fluctuation of the target nucleus or the medium in the different events of collisions. 
Correspondingly, the evaluation of a given observable should also be done by averaging over multiple different events in order to take into account such fluctuation. Note that the inclusion of coupling in \cref{eq:MV_color_charge} is a choice of convention, and no additional $g$ will appear in the interaction operator $V_{\mathcal{A}}$.

On a discrete transverse lattice, the ``uncorrelation'' defined by the 2D delta function is represented as a Kronecker delta, scaled by the lattice resolution \( a_\perp^2 \).
Similarly, in the longitudinal \( x^+ \) direction, corresponding to the background field's propagation axis, we discretize the field into \( N_\eta \) uncorrelated layers, each with a duration \( \tau \equiv L_\eta / N_\eta \).
This discretization results in the correlation function for the color charge density in~\cref{eq:MV_color_charge}, which takes the following form~\cite{Lappi:2007ku, Li:2020uhl, Li:2021zaw}:
\begin{multline}\label{eq:chgcor_dis}
  \expconfig{\rho_a(n_x,n_y,n_\tau)\rho_b({n'}_x,{n'}_y,n_\tau')}\\
 =g^2 \mu^2\delta_{ab}\frac{\delta_{n_x,{n'}_x}\delta_{n_y,{n'}_y}}{a_\perp^2}\frac{\delta_{n_\tau,n_\tau'}}{\tau}\;,
\end{multline}
where \(\vec{x}_\perp = (n_x, n_y) a_\perp\) and \(x^+ = n_\tau \tau \).
The transverse indices are \( n_x, n_y = -N_\perp, -N_\perp + 1, \ldots, N_\perp - 1 \), and the longitudinal layer indices are \( n_\tau = 1, 2, \ldots, N_\eta \).
On the amplitude level, the color charge density \(\rho\) is a stochastic variable, distributed according to a Gaussian with mean zero and variance \( g^2 \mu^2 / (a_\perp^2 \tau) \) at each lattice site \( \{ n_x, n_y, n_\tau \} \).
This distribution allows for the numerical sampling of \(\rho\) at each site and for each event.
The resulting background field can then be solved directly from~\cref{eq:poisson}.

\subsection{Time evolution of the jet}
The evolution of the jet as a quantum state is governed by the time evolution equation on the light front~\cite{Li:2020uhl},
\begin{align}
  \label{eq:ShrodingerEq}
  i\frac{\partial}{\partial x^+}\ket{\psi;x^+}=\frac{1}{2}P^-(x^+)\ket{\psi;x^+}\;.
\end{align}
 The solution of~\cref{eq:ShrodingerEq} describes the state of the investigated system at any later light-front time $x^+$,
\begin{align}\label{eq:ShrodingerEqSol}
  \ket{\psi;x^+}=\mathcal{T}_+\exp\left[-\frac{i}{2}\int_0^{x^+}\diff z^+P^-(z^+)\right]\ket{\psi;0}\;,
\end{align}
where $\mathcal{T}_+$ is the light-front time ordering.

A common approach to non-perturbative simulation of time evolution in~\cref{eq:ShrodingerEqSol} amounts to discretizing the light-front time into a large number of small steps as
\begin{align}\label{eq:time_evolution_exp}
 \begin{split}
   \mathcal{T}_+ &\exp[-\frac{i}{2}\int_0^{x^+}\diff z^+P^-(z^+)]\\
   =&\lim_{n\to\infty}\prod^{n-1}_{k=0}\mathcal{T}_+ \exp\left[-\frac{i}{2}\int_{x_{k}^+}^{x_{k+1}^+}\diff z^+P^-(z^+)\right]
 \;,
 \end{split}
\end{align}
where the step size is $\delta x^+ \equiv x^+/n$, and the intermediate times are $x_k^+=k\delta x^+ (k=0,1,2,\ldots,n)$ with $x_0^+=0$ and $x_n^+=x^+$. 
Within each short timestep, the Hamiltonian can be considered time-independent, such that the evolution operator $U_k=U(x_k^+;x_{k+1}^+)$ is written as 
\begin{align}
 \begin{split}\label{eq:trotter_1st}
   U_k\equiv &\lim_{\delta x^+\to 0}\mathcal{T}_+ \exp\left[-\frac{i}{2}\int_{x_{k}^+}^{x_{k+1}^+}\diff z^+P^-(z^+)\right]\\
  =&\exp\left[-\frac{i}{2} P^-(x_k^+) \delta x^+\right]\;.
 \end{split}
\end{align}
The evolution operator $U_k$ can be efficiently approximated using a product formula.\footnote{Note that while na\"ively in classical simulations each $U_k$ can be computed exactly by exponentiating the Hamiltonian matrix, this quickly becomes impractical as the Hilbert space size grows.
}
For instance, by partitioning the full Hamiltonian into ${P^-(x_k^+) = P_A + P_B +\cdots }$, we have
\begin{align}\label{eq:trotter}
 \begin{split}
 U_k \approx \exp\left[-\frac{i}{2} P_A \delta x^+\right]\exp\left[-\frac{i}{2} P_B\delta x^+\right] \ldots
 \;
 \end{split}
\end{align}  
In practice, the splitting of the time evolution operator into the product formula depends on the choice of the computational basis space and the details of the simulation protocol.
This allows for optimization of the numerical method based on the structure of operators in the Hamiltonian.
For example, in classical simulations performed in Ref.~\cite{Li:2021zaw}, $P_A = P^-_{\text{KE}}$ is diagonal while $P_B=V_{qg}$ is doubly block-bordered in the momentum space, whereas $P_C =V_{\mathcal{A}}$ is block-diagonal in the position space.
In comparison, on a quantum computer, the jet state is encoded in entangled qubits, and the operators are represented as quantum gates.
In this case, the choice of $P_A, P_B,\ldots$ in the product formula is associated with the encoding scheme, which we will also address in the succeeding section.

\subsection{Observables}
In the Hamiltonian formalism, observables $\hat{O}$ are represented by Hermitian operators whose measurement outcome in a given state $\ket{\psi}$ is obtained by calculating the corresponding expectation value:
\begin{align}
    \braket{\hat O} = \braket{\psi|\hat O|\psi}\;.
\end{align}
On a classical computer, the state vector is represented by its coefficients in a chosen basis (i.e., the wavefunction), while operators are represented by matrices in the same basis.
Since both are directly accessible, obtaining the expectation value is essentially a matter of matrix multiplication.
In contrast, on a quantum computer, the state vector is not directly accessible within an efficient quantum simulation of quantum many-body dynamics.
While, in principle, one could perform the \emph{full state tomography} and measure all the amplitudes describing the state of a quantum  computer, such a measurement would inevitably have cost exponential in the number of qubits.
Instead, one can map the physical observable $\hat{O}$ onto an operator acting on the multi-qubit Hilbert space (according to the chosen encoding of physical degrees of freedom), and efficiently measure the expectation value of the corresponding operator.
The starting point for such a mapping is expressing the given operator $\hat{O}$ in terms of the creation and annihilation operators acting on multi-particle states.

In studying in-medium jet evolution, we focus on observables that capture medium-induced effects, with particular emphasis on transverse momentum broadening and gluon emission.
To investigate jet broadening in transverse momentum space, we evaluate the total squared transverse momentum of the jet state at various evolution times:
\begin{align}
    \braket{\pops (x^+)} =\braket{\psi; x^+| \pops|\psi; x^+}\;.
\end{align}
The operator $\pops$, expressed in terms of quark and gluon creation and annihilation operators, is given by 
\begin{align}
\label{eq:Pperp}
\begin{split}
     \pops =& \Big(\sum_\beta b_\beta^\dagger b_\beta \vec p_{\perp} + \sum_\beta a_\beta^\dagger a_\beta \vec p_{\perp} \Big)^2
    \;,
\end{split}
\end{align}
where the summation over the mode index encompasses all basis modes in the space for respective particles. Here, $\vec{p}_\perp = (k^x, k^y)d_p$ 
and $k^j$ belong to the mode index $\beta$ [see~\cref{eq:basis_mode}].
Note that in~\cref{eq:Pperp}, we have explicitly excluded terms involving antifermions, as we have previously restricted our consideration to states of the form given in \cref{eq:jet_states}.

In the eikonal limit of $P^+=\infty$ and in the continuum, $\braket{\pops (x^+)}$ grows linear in time, and it can be derived analytically in terms of medium charge density, the IR and UV cutoffs~\cite{ Barata:2022wim, Li:2023jeh},
\begin{align}\label{eq:qhat_Eik}
  \begin{split}
  \braket{\pops (x^+)}_{\rm Eik}
  &= C_R
(g^2 \mu)^2 
 \frac{1}{4\pi}
  \Biggl\{\log\left[1+\frac{1}{(m_g a_\perp/\pi)^2}\right]\\
    & 
    -
    \frac{1}{1+(m_g a_\perp/\pi)^2}
  \Biggr\} x^+ +\braket{\pops (0)} 
    \;,
  \end{split}
\end{align}
in which $C_R$ is the eigenvalue of the $\mathrm{SU}(N_c)$ Casimir operator in the color representation of the corresponding particle.
For a single quark, $C_R = C_F = (N_c^2 - 1) / (2 N_c)$, and for a single gluon, $C_R =C_A=N_c$. For a multi-particle state, without taking into account the correlation among the particles, $C_R$ is the sum of the Casimir for each individual particle.
The neglect of particle correlations is justified for a jet initiated as a single momentum state, as in the resulting multi-particle states are not localized in the transverse position space~\cite{Li:2023jeh}.

To quantify the average number of emitted gluons, we extract the expectation value of the gluon number operator, defined as
 \begin{align}\label{eq:ng_def}
     \mathcal{N}_g\equiv \sum_\ind a^\dagger_\ind a^\pd_\ind\;,
 \end{align}
where the summation over the mode index covers all the gluon modes in the basis space.

In addition, we extract the quark (gluon) parton distribution function (PDF) for each Fock sector. 
It describes the probability of finding a quark (gluon) with longitudinal momentum fraction $z = p^+ / P^+$ in the $l$-particle sector of Fock space 
(e.g., $\ket{q}$, $\ket{qgg}$, and $\ket{gg}$),
\begin{subequations}
\begin{align}\label{eq:PDF_def_q}
    f^{(l)}_q(z) =& 
    \braket{\psi|\mathcal N^{(l)}_q(z)|\psi}
     \;,\\
    \label{eq:PDF_def_g}
    f^{(l)}_g(z) = &
    \braket{\psi|\mathcal N^{(l)}_g(z)|\psi}
    \;,
\end{align}
\end{subequations}
where the corresponding number operators can be formally written as
\begin{subequations}
\begin{alignat}{9}
    \mathcal{N}^{(l)}_q(z) = \mathcal{P}^{(l)} \mathcal{N}_q(z) \mathcal{P}^{(l)} \;,\\
    \mathcal{N}^{(l)}_g(z) = \mathcal{P}^{(l)} \mathcal{N}_g(z) \mathcal{P}^{(l)} \;.
\end{alignat}
\end{subequations}
Here, the $z$-dependent particle number operators can be written as
\begin{subequations}
\begin{align}
    &\mathcal N_q(z) =\sum_{\ind| k^+ = z K} b^\dagger_\ind b^\pd_\ind\;,
    \\
     &\mathcal N_g(z) =\sum_{\ind| k^+ = z K} a^\dagger_\ind a^\pd_\ind\;,
\end{align}
\end{subequations}
in which the summation over the mode index covers all the modes with $k^+$ satisfying the condition being specified. Note that $\mathcal N_g$ in~\cref{eq:ng_def} can be then expressed as $\sum_{k^+} \mathcal N_g(z)= \mathcal N_g$.
The $\mathcal{P}^{(l)}$ operator is the projector onto the $l$-particle sector of Fock space, and can be written in terms of operators acting on individual modes.
Such a decomposition will contain a number of terms exponential in $l$, making the usage of measurement of operators of the form $\mathcal{P}^{(l)}\hat{O}\mathcal{P}^{(l)}$ generally inefficient for large $l$. 
However, since operators $\mathcal{N}^{(l)}_q(z)$ and $\mathcal{N}^{(l)}_g(z)$ are diagonal, the numbers of particles are readily revealed from the post-measurement state.\footnote{This logic also extends to observables which can be diagonalized by particle number preserving operators.}
The full quark (gluon) PDF can then be obtained by summing over the Fock sectors $\sum_l f_q^{(l)}(z)$ [$\sum_l f_g^{(l)}(z)$].

\section{Quantum Simulation of Jet Evolution\label{sec:qsim}}

In this section, we discuss various aspects of simulating the evolution of a jet probe in medium on a digital quantum computer, including Hilbert space truncation, mapping of physical degrees of freedom onto qubits, and algorithms for state preparation and time evolution.

\subsection{Hilbert space truncation\label{ssec:input}}

In the canonical quantization of the quantum field on the light front, we have discretized the 3-dimensional momentum space, as introduced in \cref{sec:general_framework}.
The maximal value of the dimensionless momentum in the longitudinal direction is fixed by the harmonic resolution $K$ and the transverse by $N_\perp$.
Such a truncation alone is sufficient to make the Hilbert space of the jet state finite-dimensional, without additionally imposing a truncation to the Fock sector expansion; examples of the latter can be found in Refs.~\cite{Li:2020uhl, Li:2021zaw,Barata:2022wim, Barata:2023clv}.
Each single-particle mode carries some fraction of the total longitudinal momentum, denoted by $k^+$ being a positive half-integer (integer) for the boson (fermion);
in any multi-particle Fock state, all these individual momentum fractions must sum up to $K$ (see also~\cref{app:modes}).
Therefore, the maximum numbers of particles in a Fock state is upper bounded by~$\lceil K\rceil$.

Upon introducing the cutoffs, we write an arbitrary QCD Fock state of the form \cref{eq:fock_general} as
\begin{equation}
\label{eq:fock_general_modes}
\sket{
q_1^{\occf_1},
\ldots,q_{\modesq}^{\occf_{\modesq}};
\overline{q}_1^{\overline{\occf}_1},
\ldots,\overline{q}_{\modesa}^{\overline{\occf}_{\modesa}};
g_1^{\occb_1},
\ldots,g_{\modesg}^{\occb_{\modesg}}
}\;.
\end{equation}
Here, the total number of quark modes with distinct quantum numbers is denoted by 
 $\modesq$, with analogous quantities $\modesa$ for antiquarks and $\modesg$ for gluons. These numbers are generally proportional to the product of momentum space cutoffs, given by $K (2 N_\perp)^2$.
The variables $\occf_j, \overline{\occf}_j=0,1$ and $\occb_j=0,1,2,\ldots,\occmax_j$ represent the occupancies of fermionic and bosonic modes, respectively. 
The maximum occupancy of the $j$\textsuperscript{th} gluon mode, $\occmax_j$, is determined by the corresponding longitudinal momentum fraction $k^+_j$ and $K$, and is given by $\lfloor K/k^+_j\rfloor$. 
For example, given $K=3$, $\occb_j$ can take $1,2,3$ for the mode with $k^+_j=1$, but $1$ for $k^+_j=2$. 
In this manner, the finiteness of $K$ and $N_\perp$ ensures a systematic truncation of the Hilbert space, keeping it finite.

Note that for the problem of jets being investigated, we have neglected the degree of freedom of the antiquark, as indicated by  \cref{eq:jet_state_quark} and \cref{eq:jet_state_gluon}, therefore we have $\modesa=0$ for $\ket{q}_{\mathrm{jet}}$ and $\modesq=\modesa=0$ for $\ket{g}_{\mathrm{jet}}$.
The value of $K$ also indicate the maximum number of gluons that can exist simultaneously.
To further reduce the computational requirements, in our numerical simulations we will also impose a global truncation on the number of gluon modes $\occmax$ so that $\occb_j\leq\occmax$.

\subsection{Qubit encoding\label{ssec:mappings}}
\begin{figure}[t]
  \centering
  \includegraphics[width=0.46\textwidth]{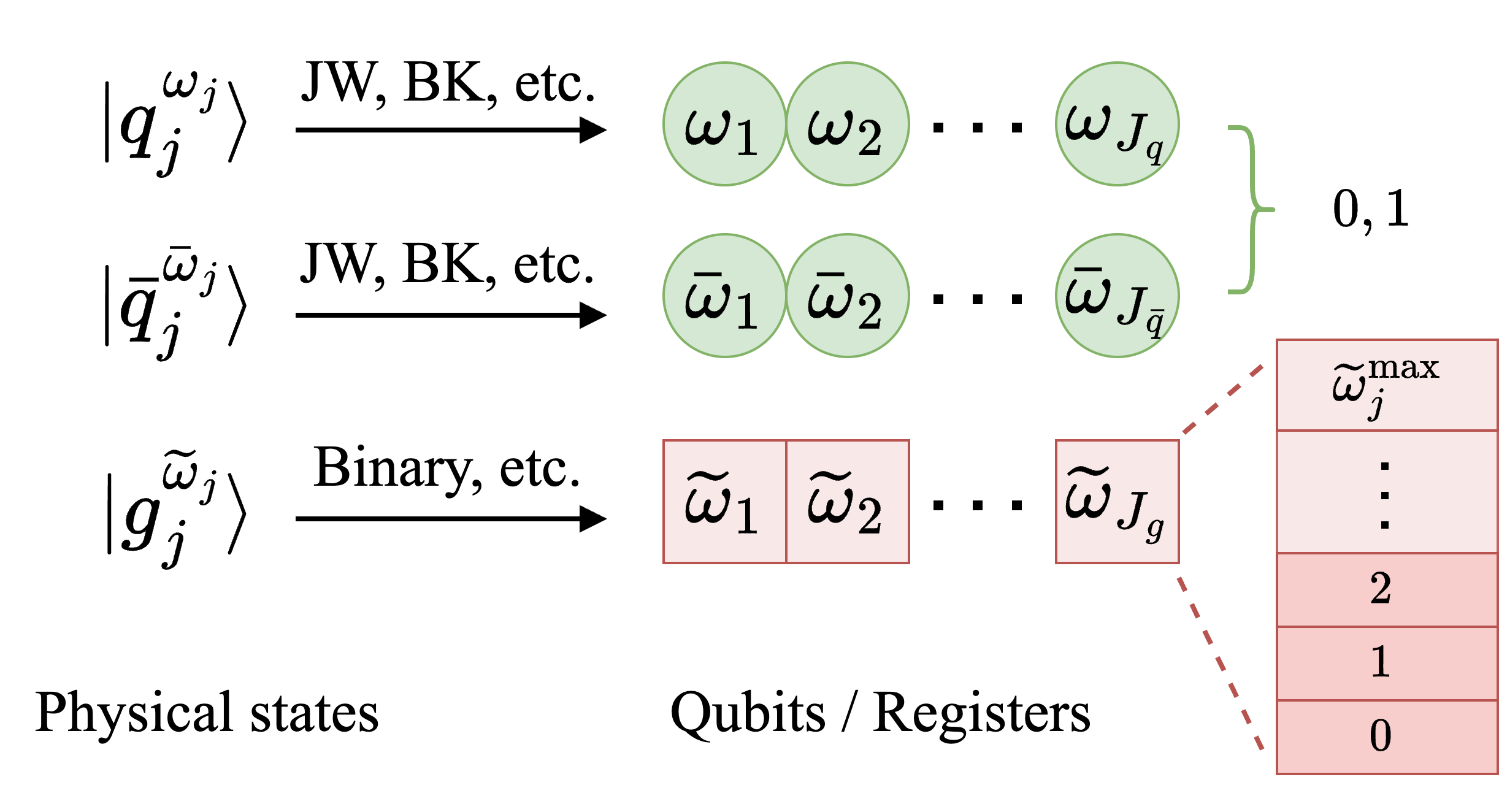}
  \caption{Schematic of the direct qubit encoding scheme, where qubit registers encode individual mode occupancies in second-quantized Fock states.
  }
 \label{fig:encoding}
\end{figure}

Qubit encoding establishes a mapping from the states of the physical model to those of a multi-qubit system.
The latter can be thought of as a system of spins which can be manipulated and measured in a highly controlled manner.
In this work, we employ the \emph{direct encoding} scheme based on the second-quantized representation of the Hamiltonian.
Here ``second-quantized'' refers to the choice of physical degrees of freedom, which in our case are the individual modes in Fock states, denoted as $q_i, \bar q_i, g_i$ in \cref{eq:fock_general_modes}.
This choice contrasts with the ``first-quantized'' representation, in which the degrees of freedom are the discretized field operators~\cite{Jordan:2011ci}.
The ``direct encoding'', in turn, refers to a situation where individual physical degrees of freedom are mapped onto fixed qubit registers.
These registers are enumerated in the same way as the single-particle states and are used to store the occupancies of corresponding modes in a Fock state, as illustrated in Fig.~\ref{fig:encoding}.

In the direct encoding scheme, the number of qubits required to represent a physical system scales linearly with the number of physical degrees of freedom (up to logarithmic factors), while the physical operators are usually represented by relatively simple operators acting on qubits.
The direct encoding schemes are to be contrasted with the \emph{compact encoding} discussed in the context of quantum chemistry~\cite{toloui2013quantum,Babbush:2017oum} and light-front quantization~\cite{Kreshchuk:2020dla,Kirby:2021ajp}, as well as with the \emph{basis state encoding} scheme used in earlier works on jet evolution~\cite{Barata:2022wim, Barata:2023clv}.
In the case of \emph{compact encoding}, one only stores the quantum numbers and occupancies of the occupied modes.
Similarly to quantum chemistry, this approach is particularly appealing in LF QCD where, due to the positivity of LF momentum, the number of particles in a Fock state cannot exceed the harmonic resolution.
While the resulting qubit Hamiltonian is not local (i.e., it contains a number of Pauli terms exponential in the number of qubits), it remains efficiently simulatable because it is sparse, enabling the use of simulation algorithms based on Hamiltonian sparsity~\cite{Kreshchuk:2020dla}.
In the case of basis state encoding, the multi-qubit basis states enumerate all the physical states one by one, which leads to a dense Hamiltonian whose matrix elements have to be determined via a classical computation, thereby making quantum simulation non-scalable.
In this work, we shall utilize the direct encoding scheme, which provides a natural starting point for exploring efficient quantum simulation algorithms.

We now discuss peculiarities of applying the direct encoding scheme to fermionic and bosonic degrees of freedom.
Due to the Pauli exclusion principle, which limits the maximum occupation of fermionic modes to 1, a finite-dimensional fermionic Fock space can be efficiently represented by a multi-qubit system, with the number of qubits matching the number of fermions.
However, one must account for the anti-commutativity of fermionic creation and annihilation operators, which can be accomplished through fermion-to-qubit mappings such as those by Jordan-Wigner (JW)~\cite{Jordan:1928wi} or Bravyi-Kitaev (BK)~\cite{Bravyi:2000vfj,seeley2012bravyi}.
A system with $\modesq$ quark and $\modesa$ antiquark modes will thus be mapped onto $\NF=\modesq+\modesa$ qubits.

The action of fermionic creation and annihilation operators on Fock states is defined via
\begin{subequations}
\label{eq:fops}
\begin{align}\label{eq:fermionic_encoding}
 &\begin{multlined}
    b^\dag_j\ket{1^{\occf_1}, \ldots, j^{\occf_j}, \ldots, N_F^{\occf_{N_F}}} = (-1)^{\sum_{j<i} \occf_j}  \\
    \times\sqrt{1-j}\ket{1^{\occf_1}, \ldots, (1-j)^{\occf_j}, \ldots, N_F^{\occf_{N_F}}},
 \end{multlined}\\
 &\begin{multlined}
    b_j\ket{1^{\occf_1}, \ldots, j^{\occf_j}, \ldots, N_F^{\occf_{N_F}}} =(-1)^{\sum_{j<i} \occf_j}  \\
    \times\sqrt{j}\ket{1^{\occf_1}, \ldots, (1-j)^{\occf_j}, \ldots, N_F^{\occf_{N_F}}}\;,
 \end{multlined}
\end{align}
\end{subequations}
where $\occf_j$ is the occupancy of the $j$\textsuperscript{th} fermionic mode and $N_F$ the total number of modes.
The fermionic operators obey the following anticommutation relations:
\begin{align}
    \{b^\dagger_j, b^\dagger_k\} = \{b_j, b_k\} = 0\;, \quad \{b_j, b^\dagger_k\} = \delta_{jk}\;.
\end{align}

For simplicity, we opt to encode the fermionic degrees of freedom using the Jordan-Wigner (JW) transformation.
In this case, while the fermionic Fock states in~\cref{eq:fops} are mapped onto multi-qubit states of exactly the same form, the creation and annihilation operators are transformed into qubit operators as follows:
\begin{subequations}
\label{eq:JW}
\begin{alignat}{9}
&b_i^\dagger &&\mapsto \big(\prod_{k=1}^{j-1}Z_{k}\big)\otimes \frac{X_j-iY_j}{2}\;,\\
&b_i &&\mapsto \big(\prod_{k=1}^{j-1}Z_{k}\big)\otimes \frac{X_j+iY_j}{2}\;.
\end{alignat}   
\end{subequations}
In the worst-case scenario, a single fermionic creation or annihilation operator is mapped onto a Pauli operator acting on \emph{all} the $\NF$ qubits, which, depending on the hardware's capabilities, may be undesirable.
This drawback of the JW transformation has motivated the development of a more efficient BK encoding, where each fermionic operator is mapped onto Pauli operators which are at most $\log N_F$-local~\cite{Bravyi:2000vfj,seeley2012bravyi}.
Note that as long as no more than one fermionic mode is occupied, the $Z$ operators do not appear on the RHS of~\cref{eq:JW}.

Bosonic operators differ from fermionic ones in that they commute with both themselves and other operators, so multiple gluons can occupy the same single bosonic mode.
Each of the $\NB=\modesg$ bosonic modes is stored in a fixed qubit register whose size typically ranges from $\lceil\log_2(\occmax_j+1)\rceil$ in binary or Gray encodings to $(\occmax_j + 1)$ in the unary encoding, with intermediate options available as well~\cite{Macridin:2018gdw, Somma:2005voa, mcardle2019digital, sawaya2019quantum, sawaya2020resource}.
The bosonic creation and annihilation operators act on the Fock states as follows:
\begin{subequations}
\begin{align}\label{eq:bosonic_encoding}
 \begin{split}
    a^\dag_j&\ket{1^{\occb_1}, \ldots, j^{\occb_j}, \ldots, N_B^{\occb_{N_B}}} \\
    &= \sqrt{\occb_{j}+1}\ket{1^{\occb_1}, \ldots, j^{\occb_j+1}, \ldots, N_B^{\occb_{N_B}}}\;,
 \end{split}\\
 \begin{split}
    a_j&\ket{1^{\occb_1}, \ldots, j^{\occb_j}, \ldots, N_B^{\occb_{N_B}}} \\
    &= \sqrt{\occb_j}\ket{1^{\occb_1}, \ldots, j^{\occb_j-1}, \ldots, N_B^{\occb_{N_B}}}\;,
 \end{split}
\end{align}
\end{subequations}
where $\occb_j$ is the occupancy of the $j$\textsuperscript{th} bosonic mode, $N_B$ is the total number of modes. 
Bosonic operators obey the following commutation relations:
\begin{align}
    [a^\dagger_j, a^\dagger_k] = [a_j, a_k] = 0\;, \quad [a_j, a^\dagger_k] = \delta_{jk}\;.
\end{align}
All the commutators between the fermionic and bosonic operator vanish.

We choose to map bosonic operators onto qubits via the standard \emph{binary encoding} which requires $\lceil\log_2(\occmax_j+1)\rceil$ qubits per mode~\cite{Macridin:2018gdw, Somma:2005voa}.
    The qubit operators acquire the following form:
\begin{subequations}
\begin{alignat}{9}
&a_j^\dagger &&\mapsto \sum_{n=0}^{\occmax_j}\sqrt{n+1}\ket{(n+1)_2}_j\bra{(n)_2}_j\;,\\
&a_j &&\mapsto \sum_{n=0}^{\occmax_j}\sqrt{n+1}\ket{(n)_2}_j\bra{(n+1)_2}_j\;,
\end{alignat}   
\end{subequations}
where $(\cdot)_2$ stands for the binary form of an integer and the $j$ subscript on the RHS indicates that the operators act on $j$\textsuperscript{th} bosonic qubit register.
In what follows, we will assume that each bosonic qubit register contains $n_b=\lceil\log_2(\occmax+1)\rceil$ qubits where $\occmax$ is a global truncation imposed on occupancies of all bosonic modes.

Using direct encoding offers several significant advantages.
Most notably, compared to the previously used basis state encoding, which involved classically precomputing the entire Hamiltonian matrix in the Fock state basis, direct encoding enables the design of algorithms where both classical and quantum resources scale polynomially with the problem size.
In a basis space specified by $N_\perp$ and $K$, we define $N_\mathrm{tot} \equiv KN^2_\perp$, which provides an estimate of the total number of modes in a Fock state. 
The fermionic and bosonic mappings introduced above allow for the physical second-quantized Hamiltonian, comprised of up to $\mathcal{O}(N_\mathrm{tot}^4)$ elementary monomials, to be mapped onto a qubit Hamiltonian containing $\mathcal{O}(N_\mathrm{tot}^4)$ Pauli operators.
This qubit Hamiltonian can then be efficiently simulated using numerous readily available algorithms, as discussed in the following sections.

\subsection{State preparation\label{ssec:stateprep}}
Depending on the physical scenario being considered, the state of the incoming quark(gluon) can either be approximated as a single momentum mode, or an eigenstate of the full vacuum Hamiltonian, before the interaction with the background field has been switched on.
In the former case, the initial state preparation is practically free in the direct encoding scheme, as it amounts to simply setting the value of the qubit register representing the mode of the incoming parton. 
In the latter case, state preparation becomes a complex problem (no easier than the simulation of time evolution discussed below) whose complexity depends on the particular technique of choice.
For example, with the usage of post-Trotter filtering-based ground state preparation techniques, the near-optimal query complexity scales as $\widetilde{\mathcal{O}}(\gamma^{-1}\Delta^{-1}||H||\log(1/\epsilon))$, where $\gamma$ is the overlap between the true ground state and the initial state, $\Delta$ is the spectral gap, ${1-\epsilon}$ is the fidelity of the final state,  $||H||$ is the Hamiltonian spectral norm, and the notation $\widetilde{\mathcal{O}}(x)$ stands for $\mathcal{O}(x\operatorname{polylog}(x))$~\cite{Lin:2020zni}. 
Here the ``query complexity'' refers to the number of calls to the \emph{block encoding} subroutine, which itself has a cost $\widetilde{\mathcal{O}}(||H||^g)$ with $g \geq 1$~\cite{Hariprakash:2023tla}. 
Near-optimal scaling with $\epsilon$ can be traded for better scaling with $||H||$ if one uses filtering algorithms based on the time evolution input model wherein, instead of utilizing the block encoding, calls to the Trotterized time evolution operator are made~\cite{Dong:2022mmq,Kane:2023jdo}.
If no prior knowledge of the initial state with a significant overlap is assumed, one can instead rely on thermalization-based algorithms~\cite{Ding:2023ytq} or use more conventional techniques such as adiabatic state preparation~\cite{farhi2000quantum,childs2001robustness,berry2020time}, along with various heuristic algorithms~\cite{peruzzo2014variational,grimsley2019adaptive}.

In this work, we study the time evolution of a jet whose state is initialized in a single-particle state (quark or gluon) of zero transverse momentum, with a fixed color.
This setup is typical when the initial jet particle forms shortly before entering the medium.
For both quark and gluon jet, this simply requires adding an \texttt{X} gate on the corresponding creation operator.
Implementing a color singlet or superposition state is also straightforward, which amounts to the preparation of W states~\cite{schon2007sequential, wang2020xy} with Hamming weight equal 1 over $N_c$ qubits for the quark jet or $N_c^2-1$ qubits for the gluon jet.

\subsection{Time evolution\label{ssec:time}}

Approaches to simulating time evolution are often categorized into two main types: those based on product formul\ae, commonly referred to as Trotter methods, and the post-Trotter methods, which typically involve constructing a \emph{Block Encoding} (BE) sub-circuit that provides access to Hamiltonian matrix elements.

Trotter methods are based on the fact that implementing a circuit to exponentiate individual Pauli terms is straightforward, while exponentiating a sum of Pauli terms can be approximated by the product of the exponents of individual terms [see also the discussion around~\cref{eq:trotter}].
Such an approximation is only valid for very small time $\delta t$ (i.e., $\delta x^+$), and to simulate evolution for longer times, the total evolution time $t$ has to be divided into many timesteps.
The gate cost of such simulations scales as $\mathcal{O}(t^{1+\frac{1}{p}}\epsilon^{-\frac{1}{p}})$ for the $p$\textsuperscript{th}-order product formula ~\cite{childs2021theory,childs2019nearly}.
The asymptotic dependence of Trotterized simulation on the problem size is highly contingent on the Hamiltonian form and typically requires a dedicated study~\cite{childs2021theory}. 
However, on general grounds, one should not expect this scaling to be better than $\mathcal{O}(||H||^{1+\frac{1}{p}})$, as this would also imply better scaling with time.\footnote{To see this, note that the total evolution time can be traded for rescaling the Hamiltonian norm: $e^{-i (\alpha t) H} = e^{-i t (\alpha H)}$.}

\begin{figure}[t]
  \centering
  \includegraphics[width=0.46\textwidth]{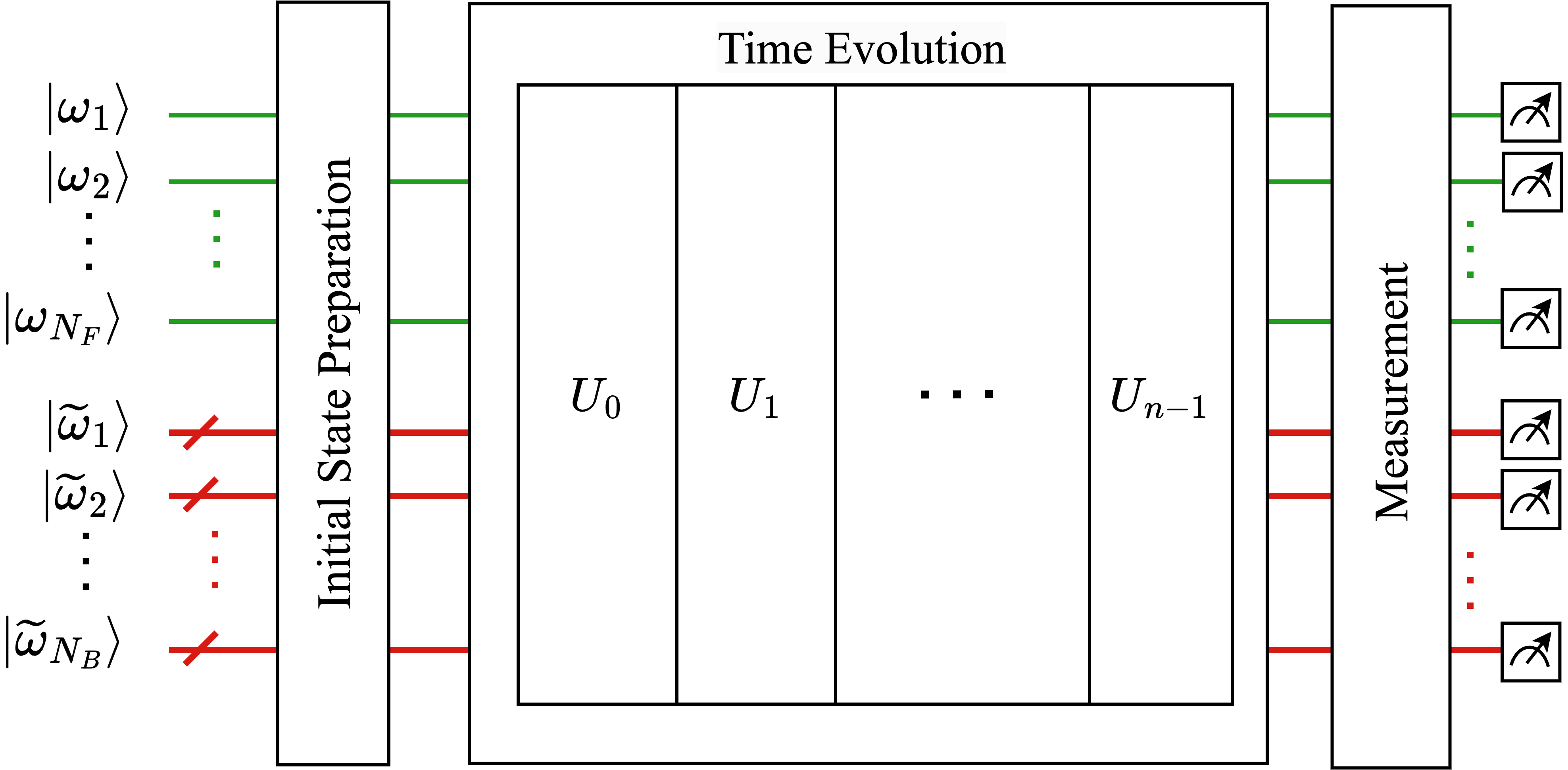}
  \caption{Schematic representation of the quantum circuit for simulating quark and gluon jets using the direct qubit encoding scheme and $n = N_t / N_\eta$ Trotter steps with $U_i$ being the evolution operator defined in \cref{eq:trotter_1st}. Green (red) lines represent the qubits encoding quark (gluon) states. A slash on qubit lines indicates multi-qubit quantum registers. The measurement circuit includes any extra gates required specifically for the measurement process, if needed.
  }
 \label{fig:circuit}
\end{figure}

Post-Trotter methods have recently been subject to active investigation in the context of simulating high-energy physics, owing to their near-optimal scaling with $t$ and $\epsilon$\cite{Kreshchuk:2020dla,Kirby:2021ajp,Hariprakash:2023tla,Kane:2024odt}, and, in certain cases, with $||H||$ and the number of colors $N_c$\cite{Rhodes:2024zbr}. 
However, comparing these methods with Trotterization is complicated by the fact that Trotterization often performs better than predicted by complexity bounds which can be notoriously challenging to calculate.
In this work, we focus on simulating time evolution via Trotterization, a conceptually simple approach that is potentially compatible with near-term devices.
We exclusively consider the first-order Trotterization method to time-evolve our light-front Hamiltonian for the jet probe in momentum space, ensuring we always use sufficiently small timesteps.
This choice is justified by examining the dependence of the observable on the number of Trotter timesteps.
We leave to future studies exploring the potential of post-Trotter simulation methods, such as those based on the Linear Combination of Unitaries~\cite{Childs:2012gwh} or Quantum Signal Processing~\cite{Low2019hamiltonian, Low2016, Martyn2021,motlagh2024generalized}.
While the direct encoding scheme is well-suited for constructing efficient block encodings~\cite{Liu:2024hmm,Du:2024zvr,Du:2024ixj,Simon:2025pbo}, we acknowledge that such implementations would be a non-trivial and interesting task.
Indeed, Trotterizing the time-dependent stochastic Hamiltonian considered in the present work is no more difficult than Trotterizing a time-independent Hamiltonian; it simply requires updating the coefficients of the Hamiltonian operator after $N_t/N_\eta$ Trotter steps, based on classically precomputed distributions, where $N_t$ is the total number of Trotter steps and $N_\eta$ is the number of layers of the background field. A schematic of the quantum circuit for simulating quark and gluon jets are presented in Fig.~\ref{fig:circuit}. 
In contrast, a post-Trotter simulation will likely necessitate either storing the Hamiltonian coefficients in a quantum database or computing them on the fly within the quantum circuit~\cite{berry2020time}.
Nevertheless, post-Trotter methods may offer cost reductions, leveraging their asymptotic optimality and the potential to utilize highly optimized Hamiltonian input schemes~\cite{Kirby:2021ajp,Du:2024ixj}.

\subsection{Measurement protocol\label{sec:meas}}

The last stage of quantum simulation is measurement, where information about specific observables, typically represented by Hermitian operators, is extracted from the final state of the system.
The optimal asymptotic cost of estimating observables to precision $h$ is given by $\mathcal{O}(1/h)$ (also known as the \emph{Heisenberg limit}~\cite{Kitaev:1995qy,Atia:2016sax,Giovannetti:2006amj,Lin:2020zni}); it can be achieved in various variants of quantum phase estimation (see Ref.~\cite{Lin:2021rwb} and references therein). 
A straightforward approach to estimating observables involves measuring individual Pauli terms in the qubit Hamiltonian~\cite{McClean:2015vup}, which has asymptotic cost $\mathcal{O}(1/h^2)$. 
As this approach is often utilized in the context of near-term simulations, such as the Variational Quantum Eigensolver (VQE), it is often referred to simply as \emph{VQE-type measurement}. 
Another near-term strategy, particularly suitable in situations where a large number of operators is estimated, is explored within the method of \emph{classical shadows}~\cite{Paini:2019sha,Huang:2020tih,Huang:2021pjy,Levy:2021dsp,Chan:2022nxf}.

In the present work, we opt to use the VQE-type measurements which acquire a particularly simple form for Hamiltonians and observables under consideration.
The second-quantized operators of all observables of our interest, such as transverse momentum broadening and longitudinal parton distribution function, can be expressed in terms of number operators of individual modes, such as $a^\dagger_\ind a^\pd_\ind$ and $b^\dagger_\ind b^\pd_\ind$.
Such operators are diagonal and, in the direct encoding scheme are represented solely by products of Pauli $Z$ operators whose expectation values are obtained by measuring the final state in the computational basis.
For instance, this implies that the large number of individual terms involved in~\cref{eq:Pperp} will not induce an overhead in the quantum measurement procedure and can be accounted for within the classical post-processing of the measurement results.

Note that, due to the stochastic nature of the background field, one additionally has to calculate configuration averages; see \cref{eq:MV_color_charge} and the related discussion.
Similar to conducting real physical experiments, this requires running calculations across various configurations of the background field, sampled from a fixed distribution, and averaging the estimated observables over these configurations.

\section{Simulation Results and Discussions\label{sec:results}}

In this section, we present the results of simulations for both quark and gluon jets using the developed framework with the {\tt Qiskit}~\cite{qiskit2024} simulator. 
Specifically, we employ the matrix product state method throughout this work for circuits with a large number of qubits (e.g., 128 qubits), which shows agreement with simulations performed using the statevector method for smaller circuits (e.g., 16 qubits) that are verified with exact diagonalization on classical computers. 

To minimize computational resource usage while preserving the physical significance, we implement two simplifications in our simulations: we take $N_c=2$ for the color and fix the light-front helicity of each single-particle mode to the up configuration. 
For the Trotterized simulations, we employ a finite timestep, which we have verified for numerical convergence by testing the results with decreasing timesteps.

\subsection{Quark jet}

We first study the quark jet scattering through a nuclear medium via the developed quantum simulation algorithm. 
Using the direct encoding scheme for the Hamiltonian operator, we successfully verified our previous results from Ref.~\cite{Barata:2022wim} for the $\ket{q}$ Fock space, as well as those in Ref.~\cite{Barata:2023clv} for the $\ket{q} + \ket{qg}$ sector. 
Additionally, we extended our simulations to include the higher $\ket{qgg}$ Fock sector. 
As mentioned earlier, we employ the Jordan-Wigner encoding~\cite{Jordan:1928wi} for fermionic operators and the standard binary encoding~\cite{sawaya2020resource} for bosonic operators. 
The total number of qubits $N_q$ required to encode a quark jet is
\begin{align}\label{eq:Nq_q_jet}
\begin{split}
    N_q &= N_F + N_B\cdot n_b\\
    &=\ceil{K} (2 N_\perp)^2 N_c  +  \floor{K} (2 N_\perp)^2 (N_c^2-1) \cdot n_b \\
    & = \mathcal{O}(K N_\perp^2 n_b)= \mathcal{O}(N_\mathrm{tot} n_b)\;,
    \end{split}
\end{align}
where $2 N_\perp$ is the number of transverse momentum modes, $K$ the resolution of longitudinal momentum modes, $N_c$ the number of colors, and $n_b=\lceil\log_2(\occmax+1)\rceil$ with $\occmax\leq \floor{K}$ the maximally allowed gluon number per mode.
Unlike the qubit cost of the encoding used in our previous works~\cite{Barata:2022wim, Barata:2023clv}, which scales linearly with the number of particles~$\occmax$ but logarithmically with the number of physical modes $N_\mathrm{tot}$, this new encoding scales logarithmically with $\occmax$ and linearly with $N_\mathrm{tot}$. 
This makes the new encoding advantageous for including more particles. 
In particular, we list the number of qubits and Pauli terms for various combinations of numerical cutoffs $N_\perp$ and $K$ in~\cref{tab:quark_jet_cost} for our simulation. 
The upper panels with $K=0.5$ show the qubit costs for simulating a quark jet in the single $\ket{q}$ Fock sector, and the lower panels with $K > 1$ for simulating a quark jet with gluon emission/absorption.

\begin{table}[b]
\setlength\tabcolsep{8pt}
 \centering
 \caption{Numbers of qubits and Pauli terms in the Hamiltonian for simulating a quark jet with SU(2) color and a fixed spin configuration.
 }
 \label{tab:quark_jet_cost} 
 \setlength\tabcolsep{6pt} 
 \begin{tabular}{ |c |  c|  c|  c| c| c| c|  c| } 
  \hline
 \multicolumn{2}{|c|}{Basis}
 & Qubits
 & \multicolumn{5}{|c|}{Hamiltonian terms}\\
 \hline
 $K$
 & $N_\perp$
 & $N_q$
 & $K_q$
 & $K_g$
 & $V_{qA}$
 & $V_{gA}$
 & $V_{qg}$
 \\
  \hline
 0.5 & 2 & 32& 30 & --& 2048& --& -- \\ 
 \hline
  0.5 & 4 & 128& 126 & --& 32768& --& -- \\ 
 \hline
 \hline
 1.5& 1 & 28& 12 & 9& 128& 96& 540 \\ 
 \hline
 2.5& 1 & 48& 18 & 18& 192& 192& 1620 \\ 
 \hline
 \end{tabular}
\end{table}

\begin{figure}[htp!]
\centering

\subfigure[\;Transverse momentum broadening\label{fig:single_quark_jet_pT}]{
\includegraphics[width=0.44\textwidth]{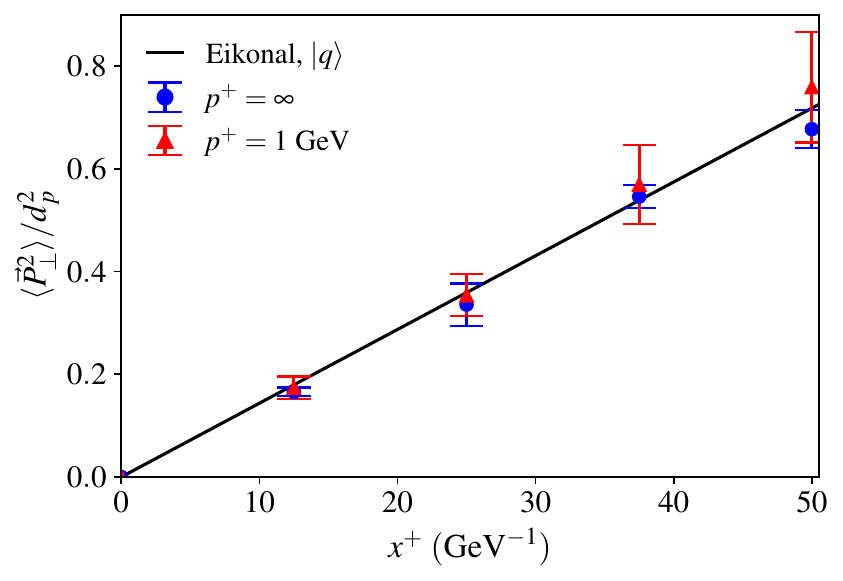}
}
\subfigure[\;Transverse probability distribution at final $x^+$\label{fig:single_quark_jet_TD}]{
\includegraphics[width=0.36\textwidth]{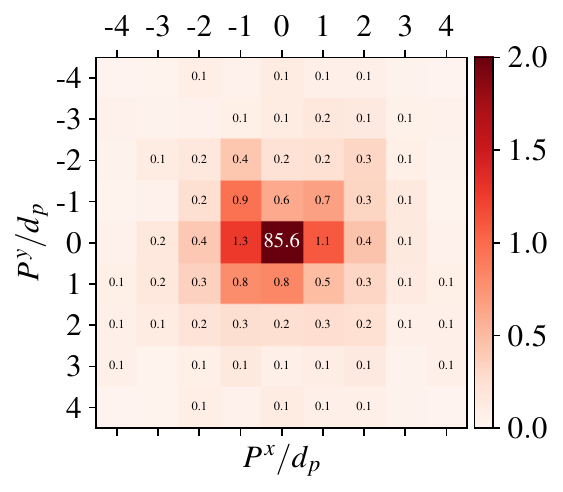}
}
\caption{Simulation results of quark jet momentum broadening in the $\ket{q}$ Fock space with $N_\perp=4$ in a medium of $g^2\mu=0.05\,\mathrm{GeV}^{3/2}$.
The solid black line is the eikonal expectation calculated according to \cref{eq:qhat_Eik}.
\label{fig:single_quark_jet}}
\end{figure}

We first consider a quark jet propagating through a medium in the leading $\ket{q}$ Fock space. 
The contributing light-front Hamiltonian can be written as
\begin{align}\label{eq:H_singleQuark}
    P^-(x^+) &= K_q + V_{qA}(x^+)\;,
\end{align}
with the quark kinetic energy term $K_q$ and the background field interaction term $V_{qA}(x^+)$, as included in~\cref{tab:second_quantized}. 
We set $m_q=0$ for the following quark jet simulations. 
Note that the coefficient for the background field is precomputed classically by solving \cref{eq:poisson} for the stochastic background field. 
The time dependence is incorporated through independent sampling at different layers along the $x^+$ direction for nuclear matter~\cite{Li:2020uhl, Barata:2022wim, Barata:2023clv}.
By minimizing the number of particles ($K=0.5$), we can achieve a relatively larger single-particle phase space ($N_\perp$) with a fixed amount of computational resources ($N_q$), thereby providing higher resolution in momentum space.
We run the simulation at $N_\perp=4$ and $L_\perp=32\; \GeV^{-1}$ using $N_q= 4 N_\perp^2 N_c =128$ qubits and 8192 shots.
The quark jet is initialized with $\vec p_\perp=\vec 0_\perp$ and either $p^+=1\;\mathrm{GeV}$ or $p^+=\infty$. In the $p^+=\infty$ case, the $K_q$ term in the Hamiltonian vanishes, reaching the eikonal limit.
For the background field, we use $N_\eta=4$ layers with medium strength of $g^2\mu=0.05$ $\mathrm{GeV}^{3/2}$. 
For the total evolution time of $L_\eta = 50 ~\GeV^{-1}$, we use a timestep of $\delta x^+=12.5\,\mathrm{GeV}^{-1}$ in the Trotterized simulation. 
From the simulation results, we investigate the momentum broadening effect and present our findings in \cref{fig:single_quark_jet}.  

In \cref{fig:single_quark_jet_pT}, we show the evolution of the transverse momentum in terms of the dimensionless quantity $ \braket{\vec P^2_\perp}/d_p^2$. The simulation results are obtained at $p^+=\infty$ and $1~\GeV$, as indicated by the plot legends. The uncertainty corresponds to the standard deviation of the expectation values obtained from $N_\mathrm{event} = 5$ configurations of the medium fields. 
The eikonal expectation shown in the solid black line is calculated according to \cref{eq:qhat_Eik}. We find that the obtained transverse momentum broadening is linear in the evolution time, which agrees well with the analytical results and also verifies the conclusions from previous jet quenching studies~\cite{Barata:2022wim} using the basis encoding.
In \cref{fig:single_quark_jet_TD}, the averaged transverse momentum probability distribution at final $x^+ = 50\;\mathrm{GeV}^{-1}$ is shown on the transverse momentum lattice, where sizable medium-induced momentum broadening is evident.

\begin{figure*}[htp!]
    \centering
    \subfigure[\;$K=1.5$, $\ket{q}+\ket{qg}$\label{fig:quark_jet_SU2_pT_K2}]{\includegraphics[width=0.47\textwidth]{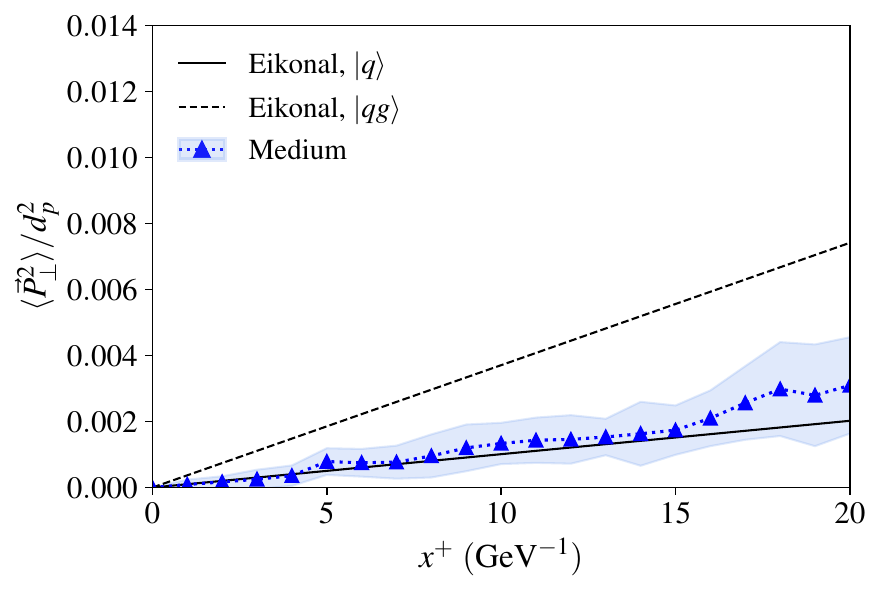}
    }\quad
    \subfigure[\;$K=2.5$, $\ket{q}+\ket{qg}+\ket{qgg}$\label{fig:quark_jet_SU2_pT_K3}]{\includegraphics[width=0.47\textwidth]{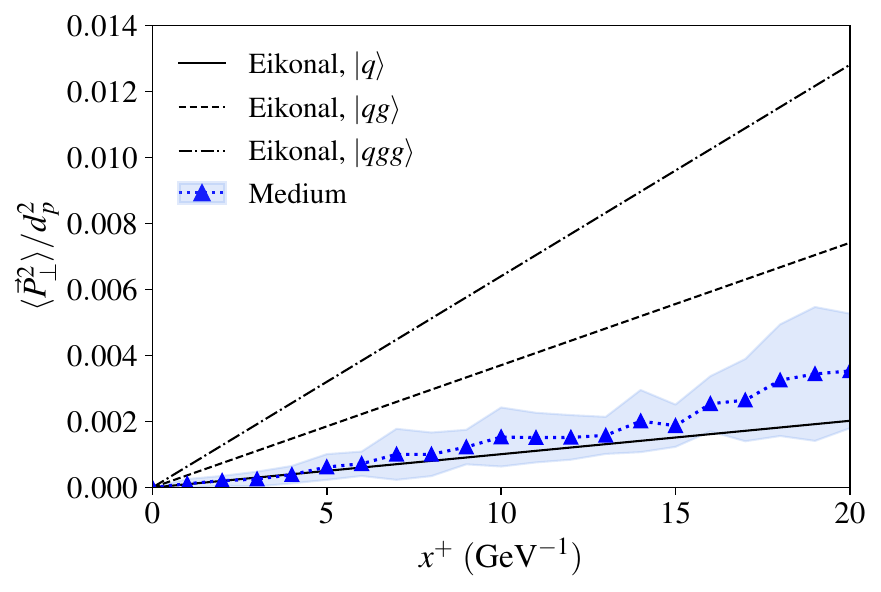}
    }
    \caption{Transverse momentum broadening obtained from quantum simulation of quark jets in the (a) $\ket{q}+\ket{qg}$ Fock space and (b) $\ket{q}+\ket{qg}+\ket{qgg}$ Fock space in a medium of $g^2\mu=0.02\,\mathrm{GeV}^{3/2}$.
    The eikonal expectations are provided according to \cref{eq:qhat_Eik} for the respective Fock sectors.
    }
\label{fig:quark_jet_pT}
\end{figure*}

\begin{figure*}[htp!]
    \centering
    \subfigure[\;$K=1.5$, $\ket{q}+\ket{qg}$\label{fig:quark_jet_SU2_number_K2}]{\includegraphics[width=0.47\textwidth]{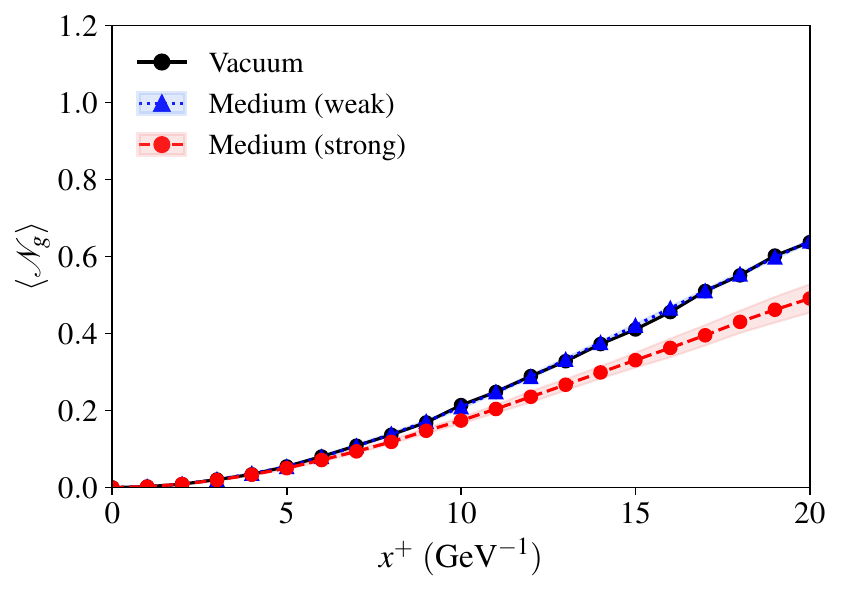}
    }\quad
    \subfigure[\;$K=2.5$, $\ket{q}+\ket{qg}+\ket{qgg}$\label{fig:quark_jet_SU2_number_K3}]{\includegraphics[width=0.47\textwidth]{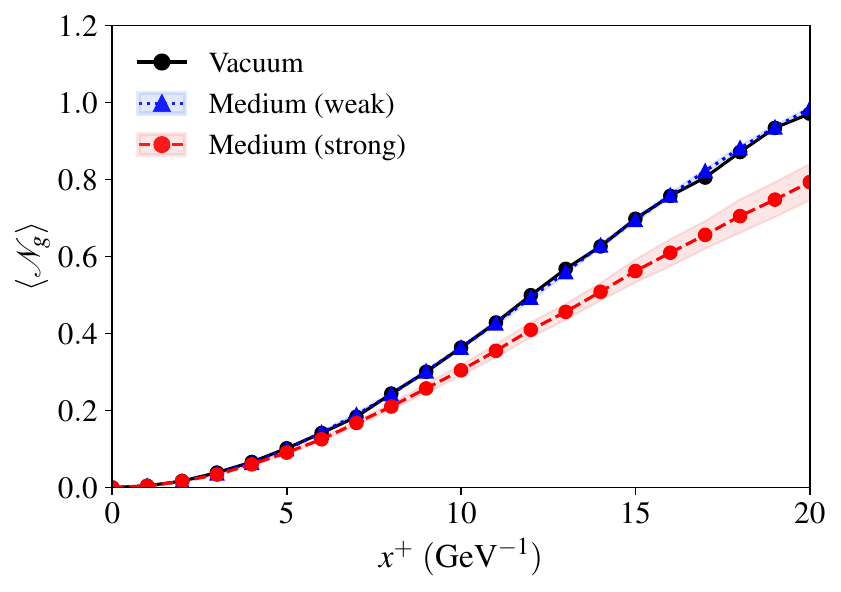}
    }
    \caption{Gluon number expectation from quantum simulation of quark jets in (a) the $\ket{q}+\ket{qg}$ Fock space and (b) the $\ket{q}+\ket{qg}+\ket{qgg}$ Fock space in vacuum, and in a weak (strong) medium of $g^2\mu=0.02\,\mathrm{GeV}^{3/2}$ ($0.5\,\mathrm{GeV}^{3/2}$). 
    }
\label{fig:quark_jet_SU2_GN}
\end{figure*}

\begin{figure*}[htp!]
    \centering
    \subfigure[\;$K=1.5$, $\ket{q}+\ket{qg}$\label{fig:quark_jet_SU2_PDF_K2}]{\includegraphics[width=0.27\textwidth]{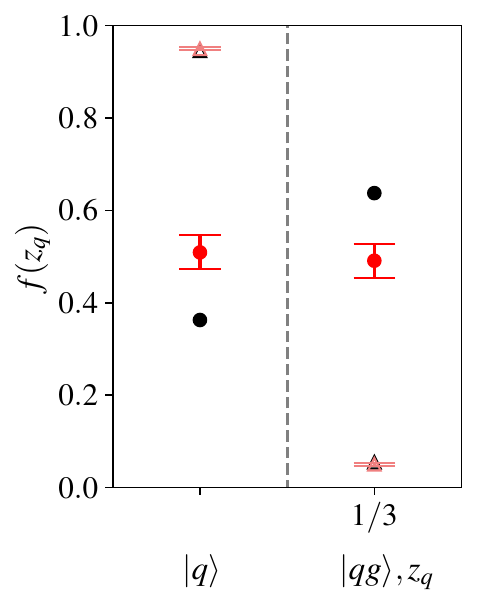}
    }\quad\quad
    \subfigure[\;$K=2.5$, $\ket{q}+\ket{qg}+\ket{qgg}$\label{fig:quark_jet_SU2_PDF_K3}]{\includegraphics[width=0.47\textwidth]{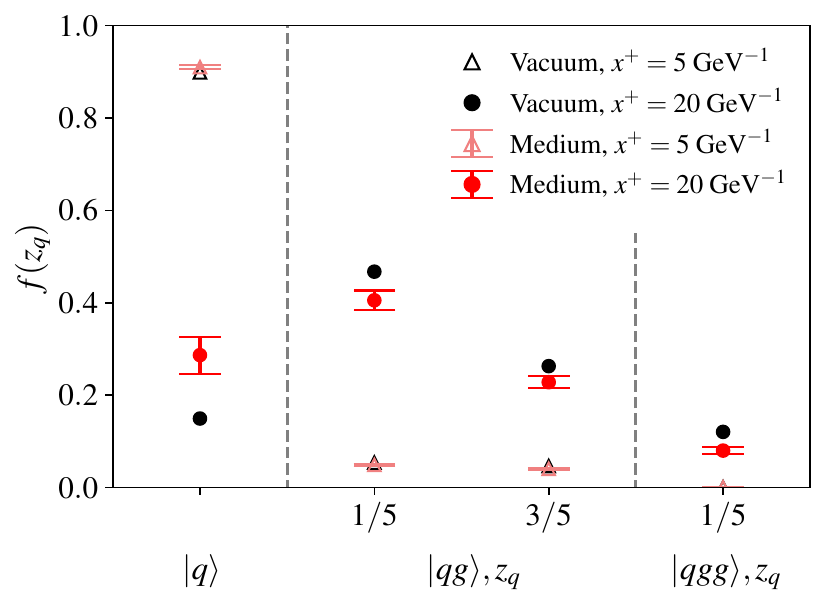}
    }
    \caption{Quark parton distribution obtained from quantum simulation of quark jets in the (a) $\ket{q}+\ket{qg}$ Fock space and (b) $\ket{q}+\ket{qg}+\ket{qgg}$ Fock space in vacuum and in a medium of $g^2\mu=0.5\,\mathrm{GeV}^{3/2}$.
    }
\label{fig:quark_jet_PDF}
\end{figure*}

Next, we examine the evolution of the quark jet with gluon emission and absorption. 
Effectively, we are working in the Fock spaces $\ket{q} + \ket{qg} + \cdots + \ket{qg \ldots g}$, where the maximum number of allowed gluons can be determined by the longitudinal momentum quanta $K$ and $\occmax$. 
We assume that the gluons are emitted solely by the quark, and not by the daughter gluons.
This allows us to simplify the quark jet Hamiltonian in~\cref{eq:Pmn_qjet} to
\begin{align}\label{eq:H_quarkJet}
    P^-(x^+) &= K_q +K_g + V_{qg}+V_{qA}(x^+) + V_{gA}(x^+)\;,
\end{align}
where $K_q, K_g$ are the kinetic energy terms, $V_{qA}(x^+), V_{gA}(x^+)$ are the background field interaction terms, and $V_{qg}$ is the gluon emission and absorption term, as included in~\cref{tab:second_quantized}.

To reduce the computational complexity of the quantum simulation, we take $n_b=1$, or equivalently $\occmax = 1$, meaning that the maximal occupancy for each gluon mode is 1. Note that we have fixed the light-front helicity of each single-particle mode to the up configuration, so only the $q(\uparrow)\leftrightarrow q (\uparrow)+g( \uparrow)$ transitions are included in $V_{qg}$.
To compare the effect of having higher Fock sectors, we simulate the quark jet evolution with both $K=1.5$ and $K=2.5$, while keeping other parameters $N_\perp=1$ and $L_\perp=32\; \mathrm{GeV}^{-1}$ the same. The key difference is that quark jets with $K=2.5$ has access to the $\ket{qgg}$ but quark jets with $K=1.5$ can only access $\ket{q}+\ket{qg}$ Fock spaces.
The simulations at $K=1.5$ and $K=2.5$ uses 28 and 48 qubits, respectively. 
In both cases, the quark jet is initialized with $\vec p_\perp=\vec 0_\perp$ and $p_q^+=P^+=1\;\mathrm{GeV}$.
To enhance the gluon emission effects, we take the coupling constant in the $V_{qg}$ term as $g_V=10$~\cite{Barata:2022wim}. 
For the background medium field, we use $N_\eta=20$ layers for the duration of the evolution $L_\eta = 20 ~\GeV^{-1}$, and we consider two different medium strengths, $g^2\mu=0.02$ $\mathrm{GeV}^{3/2}$ and $0.5$ $\mathrm{GeV}^{3/2}$.
We use a timestep of $\delta x^+=1\,\mathrm{GeV}^{-1}$ in the Trotterized simulation, for which we have verified its numerical convergence. 
The results are presented in \cref{fig:quark_jet_pT}, \cref{fig:quark_jet_SU2_GN}, and \cref{fig:quark_jet_PDF}.
All simulation results are obtained by averaging over $N_\mathrm{event}=20$ events of different medium configurations, and the resulting standard deviations represented by uncertainty bands or bars in the plots.  
For the measurement, we take 8192 shots for each configuration of the stochastic background field (which includes different realizations of $\rho_a(x^+,\bm{x})$ at timesteps).

In \cref{fig:quark_jet_pT}, we present the evolution of the transverse momentum in terms of  $ \braket{\vec P^2_\perp}/d_p^2$ for the quark jets in the $ \ket{q}+ \ket{qg}$ and $ \ket{q}+\ket{qg}+\ket{qgg}$ Fock spaces.
We run the simulation in a medium with $g^2\mu=0.02$ $\mathrm{GeV}^{3/2}$. At the corresponding saturation scale, the broadened momentum is well below the lattice UV cutoff, therefore away from lattice effect. The simulation results are shown as the blue bands, in which the triangles indicate the mean values, and the band width represents uncertainties from the configuration average.
The eikonal analytical expectations, shown in the solid, dashed, and dot-dashed lines, are obtained according to \cref{eq:qhat_Eik} with $C_R=C_F, C_F+C_A, C_F+2C_A$ for the $ \ket{q}$, $ \ket{qg}$, and $\ket{qgg}$ sectors, respectively.  
Initially, the in-medium simulation result follows the eikonal expectation of the $\ket{q}$ as the jet state is dominantly a single quark. 
Later on, the $\ket{qg}$ and $\ket{qgg} $ components emerge, increasing the rate of total momentum broadening. Comparing the momentum broadening in the $K=1.5$ and the $K=2.5$ cases, we observe a slight increase in $\braket{\vec P_\perp^2}$ in the latter, which includes contributions from the $\ket{qgg}$ sector.
The magnitude of such increase is closely related to the amount of occupation in the $\ket{qgg}$ sector but is also restricted by having limited transverse space here. 
On the lattice of $N_\perp=1$, the total momentum of the three-particle state is largely affected by periodic boundary conditions. 
Giving sufficiently large lattice size and extended evolution time, we would expect more pronounced transverse momentum broadening in cases with higher Fock sectors.

In \cref{fig:quark_jet_SU2_GN}, we present the evolution of the gluon number $\braket{\mathcal N_g}$, with the operator defined in \cref{eq:ng_def}. We run the simulations in three cases, in the vacuum, in a weaker medium with $g^2\mu = 0.02~\GeV^{3/2}$, and in a stronger medium with $g^2\mu = 0.5~\GeV^{3/2}$. 
The results in the vacuum is shown in the connected black dots, and there is no uncertainty coming from the medium configuration. The results in the two mediums are shown as the blue and red bands. 
While the result in the weaker field is similar to that in the vacuum, we observe a sizable suppression in the stronger medium for both $K=1.5$ and $K=2.5$ simulations. 
These medium-modified gluon emission effects arise from the interplay between the medium interaction and the coherent gluon emission/absorption process. Comparing the $K=1.5$ and $K=2.5$ simulation results, we see that the gluon production is significantly larger in the latter, which contains the additional $\ket{qgg}$ Fock space.

In \cref{fig:quark_jet_PDF}, we present the quark PDF for each Fock sector, as defined in \cref{eq:PDF_def_q}.
For simplicity, we omit the superscript $(l)$ in the y-axis label of the plot and explicitly indicate the Fock sector on the x-axis.
The results are taken for both the vacuum and the medium with $g^2\mu = 0.5~\GeV^{3/2}$, at an early and final times, $x^+ = 5, 20 ~\GeV^{-1}$. 
At $x^+ =5~\GeV^{-1}$, the medium results are close to the vacuum results, in both cases.
At $x^+ = 20~\GeV^{-1}$, we observe a reduction in the probability of higher Fock sectors, such as $\ket{qg}$ and $\ket{qgg}$, comparing the medium to the vacuum. This observation is consistent with the total gluon number suppression seen in \cref{fig:quark_jet_SU2_GN}. Comparing the $K=1.5$  and $K=2.5$ cases, we also observe that the probability of finding the jet in the initial $\ket{q}$ Fock sector is smaller for the latter, at both the early and final evolution times. This can be understood by noting that the inclusion of the higher $\ket{qgg}$ sector increases the chances of the state transitioning away from its initial configuration.

\subsection{Gluon jet}

To investigate the in-medium gluon jet evolution with gluon emission, we consider the Fock space of $\ket{g} + \ket{gg} + \cdots + \ket{gg\ldots g}$, where the maximum number of allowed gluons can  be determined by the resolution $K$ and, optionally, by specifying $\occmax$. 
As explained earlier, we employ the standard binary encoding~\cite{sawaya2020resource} to encode the bosonic modes, such that he total number of qubits $N_q$ required to encode a gluon jet is  
\begin{align}
\begin{split}
    N_\mathrm{q} &= N_B\cdot n_b\\
    &= 
     \floor{K} (2 N_\perp)^2 (N_c^2-1) \cdot n_b
    \\
    & = \mathcal{O}(K N_\perp^2 n_b)=\mathcal{O}(N_{\mathrm{tot}}n_b)\;,
\end{split}
\end{align}
which has the same asymptotic behavior as the quark jet, as in \cref{eq:Nq_q_jet}.
As for the quark jet, the qubit cost for the gluon jet with direct encoding grows
linearly with $n_b=\lceil\log_2(\occmax+1)\rceil$ with $\occmax\leq \floor{K}$ the maximally allowed gluon number per mode. 
The number of qubits and Pauli terms for simulating a gluon jet at two combinations of $N_\perp$ and $K$ is provided in~\cref{tab:gluon_jet_cost}.

\begin{table}[b]
\setlength\tabcolsep{8pt}
 \centering
 \caption{Numbers of qubits and Pauli terms in the Hamiltonian for simulating a gluon jet with SU(2) color and fixed spin configuration.
 }
 \label{tab:gluon_jet_cost} 
 \begin{tabular}{ |c|  c|  c|  c| c| c| c|} 
\hline
 \multicolumn{2}{|c|}{Basis}
 & Qubits
 & \multicolumn{4}{|c|}{Hamiltonian terms}\\
 \hline
 $K$
 & $N_\perp$
 & $N_q$
 & $K_g$
 & $V_{gA}$
 & $V_{ggg}$ 
 & $V_{gggg}$
 \\
 \hline
 2& 1 & 24& 18 & 192& 144& 2304\\ 
 \hline
 3& 1 & 36& 27 & 288& 504& 10944\\ 
 \hline
 \end{tabular}
\end{table}

To simplify the computation, we turn off the four-gluon and instantaneous interactions, thereby reducing the gluon jet Hamiltonian in~\cref{eq:Pmn_gjet} to 
\begin{align}\label{eq:H_gluonJet}
   P^-(x^+) &= K_g + V_{gA}(x^+) +  V_{ggg} \;,
\end{align}
where $K_g$ is the gluon kinetic energy, $V_{gA}(x^+)$ is the background field interaction, and $V_{ggg}$ is the three-gluon interaction, all as included in~\cref{tab:second_quantized}. 
As in the study of the quark jet, we also take the coupling constant in the $V_{ggg}$ term as $g_V=10$ to obtain a sizable gluon emission. 
Likewise, we set $\occmax = 1$ for simulating the gluon jet in order to reduce the computational complexity of the quantum simulation.

Note that specially at $K=2$ and $N_\perp=1$, the transition from a single gluon with $0$ transverse momentum to two gluons vanishes. Consider the transition $g(k_1, a_1)\to g(k_2, a_2) + g(k_3, a_3)$, in which $k_1^+=2$ and $k_1^x=k_1^y=0$, then by momentum conservation, the two final gluons on such a small basis have to have the same three momentum, $ k_2^+=k_3^+=1$ and $ (k_2^x, k_2^y) =(k_3^x, k_3^y)= (0,0), (-1,0), (0,-1), (-1,-1)$, taking into account the periodic boundary conditions. Then the two final gluons only differ in color, $a_2\neq a_3$. Since the color factor in the transition operator $f^{a_1 a_2 a_3}$ is completely antisymmetric, the two transition amplitudes by exchanging $a_2\leftrightarrow a_3$ cancels out. As a result, no gluon emission or absorption happens there. 
Though one could alternatively start with an initial $k_\perp \neq 0$ at $K=2$ and $N_\perp=1$ to bypass this special vanishing mechanism (e.g., the implementation in ~\cite{Yao:2022eqm}), we maintain consistency in our study by always beginning with a $k_\perp = 0$ state and will not consider this case. Note that at a larger basis space, with either increased $K$ or $N_\perp$, the two final gluons no longer need to occupy the same momentum, and the transition survives. 
In the following, we focus on $K=3$ with the $\ket{g} + \ket{gg} + \ket{ggg}$ Fock space.

\begin{figure}[thp!]
    \centering
    \subfigure[\;Momentum broadening\label{fig:gluon_jet_SU2_pT}]{\includegraphics[width=0.46\textwidth]{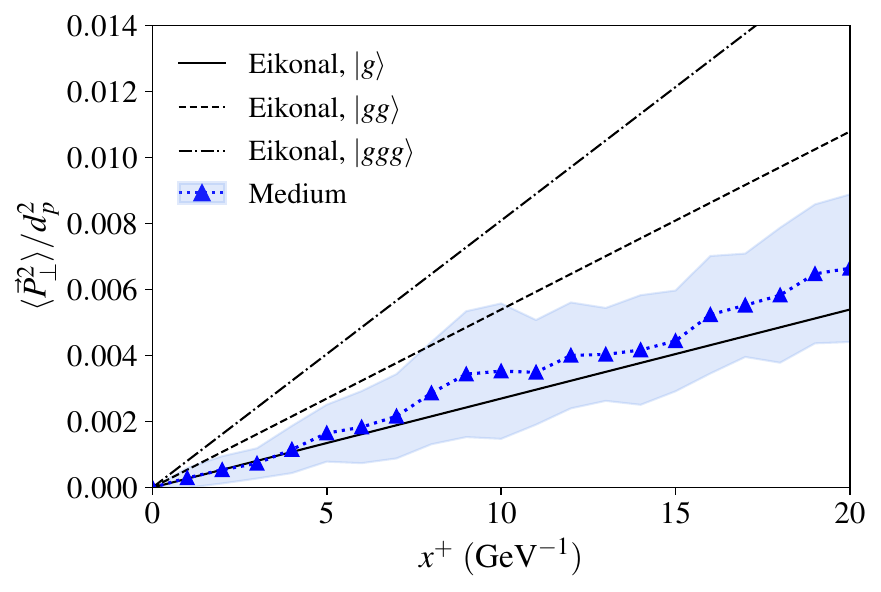}
    }
    \subfigure[\;Gluon number\label{fig:gluon_jet_SU2_ng}]{\includegraphics[width=0.45\textwidth]{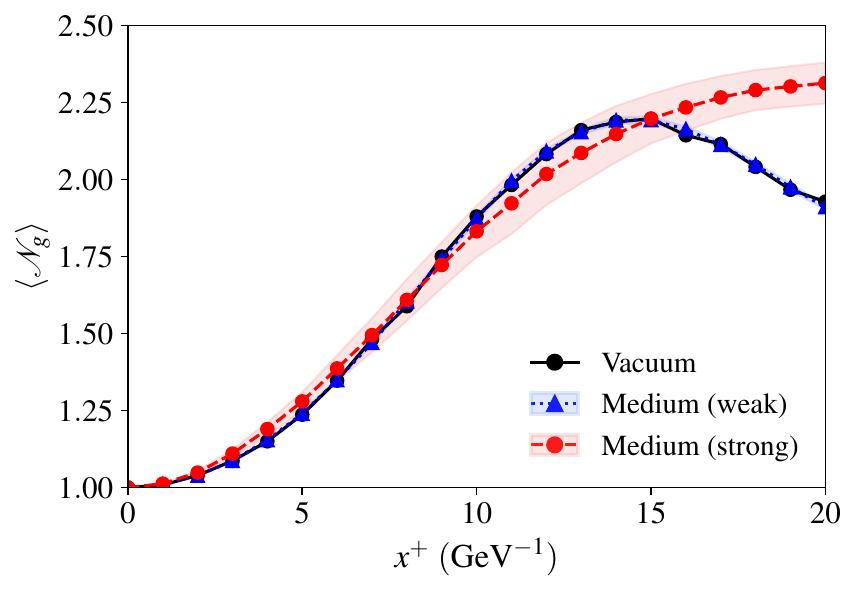}}
    \subfigure[\;Gluon parton distribution function\label{fig:gluon_jet_SU2_PDF}]{\includegraphics[width=0.44\textwidth]{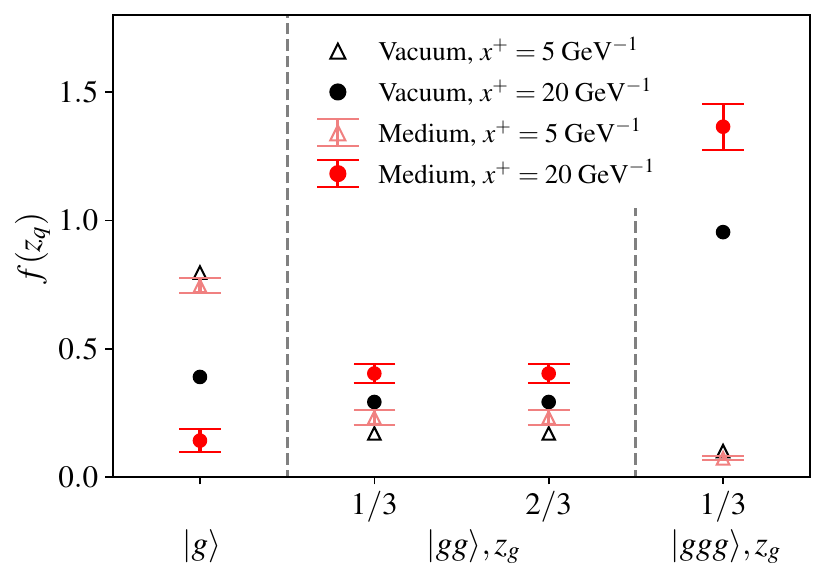}
    }
    \caption{Quantum simulation of gluon jet with SU(2) color with $K=3$ longitudinal modes in vacuum and in a weak (strong) medium of $g^2\mu=0.02\,\mathrm{GeV}^{3/2}$ ($0.5\,\mathrm{GeV}^{3/2}$). 
    }
\label{fig:gluon_jet_SU2}
\end{figure}

We simulate the gluon jet evolution in the space specified by $K=3$, $N_\perp=1$ and $L_\perp=32\; \mathrm{GeV}^{-1}$, for which we use 36 qubits on the quantum simulator.
The gluon jet is initialized with $\vec p_\perp=\vec 0_\perp$ and $p_g^+=P^+=2\pi K/L=1\;\mathrm{GeV}$.
As in the study of the quark jet, we set $N_\eta=20$ layers of the medium in the duration of $L_\eta = 20 ~\GeV^{-1}$, and we consider two different medium strengths, $g^2\mu=0.02$ $\mathrm{GeV}^{3/2}$ and $0.5$ $\mathrm{GeV}^{3/2}$. The timestep is set to $\delta x^+=1\,\mathrm{GeV}^{-1}$.
All simulation results are averaged over $N_\mathrm{event}=20$ events of different medium configurations, and we perform 8192 shots for the measurement of each configuration.
The results are presented in \cref{fig:gluon_jet_SU2}.

In \cref{fig:gluon_jet_SU2_pT}, we present the evolution of the transverse momentum in terms of  $ \braket{\vec P^2_\perp}/d_p^2$ for the gluon jet in a medium with $g^2\mu=0.02$ $\mathrm{GeV}^{3/2}$. 
The eikonal analytical expectations, shown in the solid, dashed, and dot-dashed lines, are obtained according to \cref{eq:qhat_Eik} with $C_R=C_A, 2C_A, 3C_A$ for the $ \ket{g}$, $ \ket{gg}$, and $\ket{ggg}$ sectors, respectively.  
The simulation result, shown as the blue band, initially follows the eikonal expectation of the $\ket{g}$, and later on, exhibits a faster rate of increase due to contributions from the emerging higher Fock sectors.
This observation is similar to what we have seen in the quark jet. 
The evolution of the gluon number can facilitate the understanding of the aforementioned findings. 

In \cref{fig:gluon_jet_SU2_ng}, we show the evolution of the gluon number $\braket{\mathcal{N}_g}$, with the operator defined in \cref{eq:ng_def}. 
We run the simulations in three cases, in the vacuum, in a weaker medium with $g^2\mu = 0.02~\GeV^{3/2}$, and in a stronger medium with $g^2\mu = 0.5~\GeV^{3/2}$. 
The result in the vacuum is shown in the connected black dots, and the results in the two mediums are shown as the blue and red bands. 
Note that for the Fock space being considered, $1 < \braket{\mathcal{N}_g} < 3$. 
In all three cases presented, the value of $\braket{\mathcal{N}_g}$ is initially 1, then gradually grows, showing a trend of oscillation over time. 
In the presence of the weaker medium, the gluon number does not significantly differ from that in the vacuum, indicating negligible medium modification. 
In contrast, in the stronger medium, the gluon number exhibits a sizable difference compared to the vacuum case. 
The medium modification switches several times between suppression and enhancement.
This behavior suggests non-trivial interference effects between the medium interaction and the coherent gluon emission/absorption, especially in the earlier stages.
In the later stage, as the different multi-gluon states have fully decohered in terms of their relative phases, the additional modification from the medium is less pronounced. Instead, as the medium interaction opens up a larger phase space for the transitions into the multi-gluon states (e.g., a two-gluon state in the color-singlet configuration), the total gluon emission probability is more likely to get enhanced. 
A more systematic study is required to achieve a comprehensive understanding of the underlying mechanism.

Lastly, in \cref{fig:gluon_jet_SU2_PDF}, we present the gluon PDF $f(z_g)$ as a function of the gluon's longitudinal momentum fraction $z_g=p_g^+/P^+$ for each Fock sector. The definition is provided in \cref{eq:PDF_def_g}. 
The results are taken for both the vacuum and the medium with $g^2\mu = 0.5~\GeV^{3/2}$, at an early and final times, $x^+ = 5, 20 ~\GeV^{-1}$. 
At $x^+ =5~\GeV^{-1}$, the medium results are close to the vacuum results, but with a larger difference compared to that in the quark jet, as in \cref{fig:quark_jet_PDF}.
At $x^+ = 20~\GeV^{-1}$, we observe an enhancement in the probability of higher Fock sectors, $\ket{gg}$ and $\ket{ggg}$, comparing the medium to the vacuum. 
This finding is consistent with the final gluon number enhancement seen in \cref{fig:quark_jet_SU2_GN}. 
In addition, the gluons with $z = 1/3$ in the $\ket{ggg}$ Fock sector tend to dominate at the final time, and the occupancy in the initial $\ket{g}$ sector becomes small. This indicates the importance of including higher Fock sectors to accurately capture gluon emission effects.

\section{Conclusion and Outlook}\label{sec:summary}

In this work, we described a unified and scalable approach to the quantum simulation of the high-energy phenomena associated with the evolution of quantum jets in a medium, using the non-perturbative light-front Hamiltonian approach.
Our method features the construction of second-quantized light-front Hamiltonians and their subsequent mapping onto qubits through a direct encoding scheme, where qubit degrees of freedom store the occupancies of second-quantized fermionic~\cite{Jordan:1928wi,Bravyi:2000vfj,seeley2012bravyi} and bosonic modes~\cite{Macridin:2018gdw, Somma:2005voa}.
As compared to the basis encoding scheme used in previous investigations~\cite{Barata:2022wim, Barata:2023clv}, direct encoding leads to a quantum algorithm whose cost (number of qubits, gate depth) scales polynomially with the system size (cutoffs on single-particle degrees of freedom, number of particles in a Fock state).
These advantages become increasingly significant with a higher number of Fock particles for the jet, making the approach a natural step toward reaching quantum advantage.

To demonstrate the feasibility of our methods, we simulate the in-medium evolution of quark and gluon jets up to three leading Fock sectors: $\ket{q} + \ket{qg} + \ket{qgg}$ and $\ket{g} + \ket{gg} + \ket{ggg}$, respectively. Despite several simplifying assumptions (e.g., SU(2) color) and limitations on lattices, we present the first calculations of quark and gluon jets with up to three particles.
This quantum simulation, based on the usage of Trotter product formula and performed on a classical emulator, surpasses existing classical simulations.
With our quantum simulation algorithm, we can directly access real-time observables in the sub-eikonal region, allowing us to observe medium modifications to the jets.
Specifically, we investigate transverse momentum broadening, average gluon number production, and the jet parton probability distribution for both jets, comparing our findings with eikonal expectations.

Our simulation strategy is an important first step towards simulating jet physics on quantum computers.
Based on the usage of direct encoding scheme, such a simulation would naturally incorporate Focks states with a large number of particles, with minimum changes in the algorithm cost.
A realistic treatment of quark and gluon jets requires accounting for instantaneous interactions and many-particle states.
In classical simulations, the latter poses a significant limitation, as including sectors with a larger number of particles rapidly increases memory demands.
Conversely, a quantum device naturally encodes many-particle quantum states, yet it faces a more critical constraint in the maximum circuit depth, which is defined by the complexity of the Hamiltonian, i.e., by the interactions considered.

The quantum simulation algorithm can be enhanced in several ways.
Since medium interaction terms and kinetic energy terms are diagonal in the position and momentum bases, respectively, one could leverage the ability to switch between these bases on the fly using the quantum Fourier transform circuit---an approach explored in both Trotter~\cite{Jordan:2011ci,Klco:2018zqz,Barata:2023clv} and post-Trotter~\cite{Hariprakash:2023tla} methods for time evolution.
Alternating simulation in momentum and position spaces related by quantum Fourier transform is likely to reduce the gate complexity but may require implementation of quantum Fourier transform for unary-encoded states~\cite{jain2024quantum} and usage of qudits operations~\cite{Pavlidis2021} in order to deal with multiple bosonic excitations.
As simulations scale to larger system sizes, it may become necessary to carefully compare the simulation costs of Trotter and post-Trotter methods, as recent studies in equal-time lattice formulations have done~\cite{Hariprakash:2023tla,Rhodes:2024zbr,Kane:2024odt}.
Additionally, exploring alternative encoding schemes~\cite{Kirby:2021ajp,Du:2024ixj} could reveal further cost reductions.
A different and equally interesting problem is to study the time evolution of an incoming dressed quark state \cite{Li:dressedquark}, which can be prepared on the circuit as the eigenstate of the QCD Hamiltonian using filter-based state preparation techniques.

\section*{Acknowledgement}
We are grateful to João Barata, Weijie Du, James P. Vary, Bin Wu, Siqi Xu, and Xingbo Zhao for their helpful and valuable discussions. 
This work is supported by the European Research Council under project ERC-2018-ADG-835105 YoctoLHC; by Maria de Maeztu excellence unit grant CEX2023-001318-M and project PID2020-119632GB-I00 funded by MICIU/AEI/10.13039/501100011033; and by ERDF/EU. It has received funding from Xunta de Galicia (CIGUS Network of Research Centres). WQ is also supported by the Marie Sklodowska-Curie Actions Postdoctoral Fellowships under Grant No. 101109293.
MK was supported by the DOE, Office of Science under contract DE-AC02-05CH11231, partially through Quantum Information Science Enabled Discovery (QuantISED) for High Energy Physics (KA2401032). 

\appendix

\section{Conventions}

Throughout this paper, we follow the conventions in Refs.~\cite{Li:2021zaw, Barata:2023clv} and provide additional identities relevant to this work.

\label{app:convention}
The light-front coordinates are defined as \( (x^+, x^-, x^1, x^2) \), where \( x^+ = x^0 + x^3 \) represents the light-front time, \( x^- = x^0 - x^3 \) the longitudinal coordinate, and \( \vec{x}_\perp = (x^1, x^2) \) the transverse coordinates.
For convenience, we denote the transverse indices by \( i = 1, 2 \) or \( i = x, y \) interchangeably.
Similarly, the four-momentum vector is given by \( (p^+, p^-, p^1, p^2) \), where \( p^+ = p^0 + p^3 \) denotes the longitudinal momentum, \( p^- = p^0 - p^3 \) the light-front energy, and \( \vec{p}_\perp = (p^1, p^2)= (p^x, p^y) \) the transverse momenta. The momentum square is denoted by $ p^2_\perp = (p^x)^2+ (p^y)^2$.

The non-vanishing elements of the metric tensor are given by
\begin{align}
\begin{split}
 & g^{+-} = g^{-+} = 2\;, \quad g_{+-} = g_{-+} = \frac{1}{2}\;, \\
 & g^{11} = g^{22} = -1\;.
\end{split}
\end{align}

For any transverse two-dimensional vector \( \vec{n}_\perp = (n_x, n_y) \), we define \( n^R \equiv n_x + i n_y \) and \( n^L \equiv n_x - i n_y \).

 \section{Quantization of the light-front QCD Hamiltonian in a discretized momentum space}
\label{app:Pmin}

In this appendix, we outline the details of our quantization scheme and present the Hamiltonian in its second-quantized form.
Light-front quantization of QCD in a discretized momentum representation, also known as Discretized Light-Cone Quantization (DLCQ), was originally introduced in Refs.~\cite{Pauli:1985pv,Pauli:1985ps,Eller:1986nt,Harindranath:1987db,Hornbostel:1988fb} for calculations of relativistic bound-state calculations (see Chapter 4 of Ref.\cite{Brodsky:1997de} for a review).
It was later adapted to tackle time-dependent problems in the presence of an external background field within the time-dependent Basis Light-Front Quantization (tBLFQ) method~\cite{Li:2020uhl, Li:2021zaw, Li:2023jeh}, and subsequently extended to quantum simulation applications in Refs.~\cite{Barata:2022wim, Barata:2023clv} and in this work.

\subsection{Quantization in a discrete space}\label{app:modes}

We consider a system contained within a box of finite volume \(\Omega = 2L \Omega_\perp\), where \(\Omega_\perp = (2L_\perp)^2\). We have introduced two artificial length parameters: \(L\) in the longitudinal direction and \(L_\perp\) in the transverse directions. In the longitudinal direction, where \(-L \leq x^- \leq L\), we impose periodic boundary conditions for bosons and anti-periodic boundary conditions for fermions. This leads to a discretization of the longitudinal momentum space given by:
  \begin{align}
      &p^+ = 
      \begin{cases}
          k^+ d_+, ~\text{with}~ k^+ = \frac{1}{2}, \frac{3}{2}, \ldots,\infty 
          ~\text{for fermions}\;, \\
          k^+ d_+, ~\text{with}~ k^+ = 1,2, \ldots,\infty
          ~\text{for bosons}\;,
      \end{cases}
  \end{align}
where the unit of $p^+$ is $d_+\equiv 2\pi/L$.

Consider a quantum system with a definite total longitudinal momentum \(P^+\), corresponding to the total dimensionless momentum \(K\), such that \(P^+ = K d_+\)~\cite{Pauli:1985pv}.
Each multi-particle state of the form~\cref{eq:fock_general_modes} then satisfies
\begin{equation}
    \sum_{j=1}^{\modesq} \occf^\pd_j k^+_j
    +
    \sum_{j=1}^{\modesa} \overline{\occf}^\pd_j k^+_j
    +
    \sum_{j=1}^{\modesg} \occb^\pd_j k^+_j
    =
    K \;,
\end{equation}
where $k^+_j$ is the longitudinal momentum of the $j$\textsuperscript{th} quark, antiquark, or gluon mode.

In the transverse dimension,  $-L_\perp \le r^i\le L_\perp$, we impose the periodic boundary conditions and discretize the space into a square lattice,
  \begin{align}
    r^i= n_i a_\perp \quad (i = x,y)  \;,
 \end{align}
where $n_i=-N,-N+1,\ldots,N-1 $ and the lattice spacing is $a=L_\perp/N_\perp$. 
  The corresponding momentum space is also discrete with periodic boundary conditions,
  \begin{align}
     & p^i = k^i d_p \;,
  \end{align}
where $k^i = -N_\perp, -N_\perp+1 \ldots, N_\perp-1$ and $d_p\equiv \pi/L_\perp$ is the resolution in momentum space. 
    
The mode expansion of the field operators in such a discrete momentum basis is
\begin{subequations}
\label{eq:box_modes}
    \begin{align}
    \begin{split}
    \Psi_{\mu,c} (x)=  &\sum_{\bar{\alpha}} \frac{1}{\sqrt{ p^+ 2L (2L_\perp)^2}}
     [b^\pd_{\ind} u_\mu(p,\lambda)
       e^{-ip\cdot x}\\
       &+d^\dagger_{\ind} v_\mu(p,\lambda) e^{ip\cdot x}]\;, 
    \end{split}
    \\
     \begin{split}
     A_{\mu,a}(x)=  &\sum_{\bar{\alpha}} \frac{1}{\sqrt{ p^+ 2L (2L_\perp)^2}}
      [a^\pd_{\ind}\epsilon_\mu(p,\lambda)e^{-ip\cdot x}\\
       &+a^\dagger_{\ind}\epsilon_\mu^*(p,\lambda)e^{ip\cdot x}]   \;,  
       \end{split}
      \end{align}
  \end{subequations}
    where $p\cdot x=1/2p^+ x^- -\vec p_\perp\cdot \vec x_\perp$ is the 3-product for the spatial components of $p^\mu$ and $x^\mu$.
  Each single particle state is specified by five quantum numbers, 
  \begin{align}
      \ind=\{ k^+, k^x, k^y, \lambda,c(a) \}\;,
  \end{align}
   where $\lambda$ is the light-front helicity, and $c=1,2,\ldots, N_c$ ($a=1,2,\ldots N_c^2-1 $) is the color index of the quark/antiquark (gauge) field.
   For the convenience, we also define $\bar{\alpha}=\{ k^+, k^x, k^y, \lambda \}$ and note that $\ind = \{\bar{\alpha},c(a)\}$.
  The creation operators $b^{\dagger}_\ind$, $d^{\dagger}_\ind$ and $a^{\dagger}_\ind$
  create quarks, antiquarks and gluons with quantum numbers $\ind$ respectively. 
  They obey the following commutation and anti-commutation relations,
  \begin{align}\label{eq:commutations}
      \begin{split}
   & \{b^\pd_\ind,b^{\dagger}_{\ind'}\}=\{d^\pd_{\ind},d_{\ind'}^{\dagger}\}
  =[a^\pd_{\ind},a^{\dagger}_{\ind'}]=\delta^\pd_{\ind,\ind'} \;,  
      \end{split}
  \end{align}
  with all the other commutators vanishing.
  
The fields obey the standard equal-light-front-time commutation relations~\cite{Brodsky:1991ir}, which for dynamical fields acquire the form of
  \begin{align}
    \begin{split}  
      \{\Psi_{+,c}(x), \Psi^{\dagger }_{+,c'}(y)\}_{x^+=y^+}
       =& \\
       \Lambda_+ \delta(x^- - y^-) &\delta^2(\vec x_\perp - \vec y_\perp)\delta_{c,c'}\;, 
    \end{split}
  \end{align}
  in which $ \Lambda_+ =\gamma^0\gamma^+/2$ is the light-front projector, 
  and
\begin{align}
    \begin{split}
      [A_{i, a}(x), A^{\dagger }_{j,b}(y)]_{x^+=y^+}
     =& \\-\frac{i}{4} \epsilon(x^- - y^-) &\delta^2(\vec x_\perp - \vec y_\perp)\delta_{i,j} \delta_{a,b}
     \;,
    \end{split}
  \end{align}
with $i,j=1,2$, and $\epsilon(x)$ is the sign function.

The single-particle basis states are defined as 
  \begin{align}
  \begin{split}
       &\ket{q(\ind)}
    =b^\dagger_{\ind}\ket{0}\;,\\
           &\ket{\bar q(\ind)}
    =d^\dagger_{\ind}\ket{0}\;,\\
           &\ket{g(\ind)}
    =a^\dagger_{\ind}\ket{0}\;.
  \end{split}
  \end{align}
The multiple-particle basis states can be constructed as tensor products of the above states.  

\subsection{The light-front QCD Hamiltonian with background field}
\label{app:Pmin_2nd}

The QCD light-front Hamiltonian with a background field \(\mathcal{A}^\mu_a\) can be obtained from the QCD Lagrangian by replacing \(A^\mu_a\) with \(A^\mu_a + \mathcal{A}^\mu_a\)~\cite{Li:2020uhl}\footnote{Note that the background field must be in the light-cone gauge, specifically \(\mathcal{A}^+ = A^+ = 0\).}:
\begin{align}
    P^-(x^+) = P^-_{QCD} + V_{\mathcal{A}}(x^+)\;.
\end{align}
Here, $P^-_{QCD}$ is the \emph{vacuum QCD Hamiltonian}, i.e., the standard QCD Hamiltonian without the background field~\cite{Brodsky:1997de}, while $V_{\mathcal{A}}(x^+)$ contains the interactions with the background field $ \mathcal A$. 
The vacuum Hamiltonian can be written as
\begin{align}
    \label{eq:PQCD_app}
     P^-_{QCD} =P^-_{\text{KE}} + V_{qg}+V_{ggg}+V_{gggg}+W_{g} + W_{f}\;.
\end{align}
The first term in~\cref{eq:PQCD_app} is the kinetic energy operator,
\begin{align}\label{eq:PKE}
  \begin{split}
    P^-_{\text{KE}}=&\int\diff x^-\diff^2 x_\perp
    \bigg\{
    -\frac{1}{2}A^j_a{(i\nabla)}^2_\perp A_j^a\\
    &
    +\frac{1}{2}\bar{\Psi}\gamma^+\frac{m^2-\nabla_\perp^2}{2i\partial_-}\Psi
    \bigg\}
    \;.
  \end{split}
\end{align}
The three-point-gluon vertex is given by
\begin{align}\label{eq:LFH_Vggg}
  \begin{split}
    V_{ggg}=&\int\diff x^-\diff^2 x_\perp
    g  f^{abc}\partial^\mu A^i_a A_i^b A^c_\mu 
 \;.
  \end{split}
 \end{align}
The four-point-gluon vertex is given by
\begin{align}
\begin{split}
 V_{gggg}=&\int\diff x^-\diff^2 x_\perp
 \frac{g^2}{4}  f^{abc} A^i_b A^j_c 
 f^{aef} A_i^e A_j^f
\;.
\end{split}
\end{align}
The instantaneous gluon vertex is given by
\begin{align}
\begin{split}
  W_{g}=&\frac{1}{2}g^2
   \int\diff x^-\diff^2 x_\perp
   J^+_a\frac{1}{{(i\partial^+)}^2}J^+_a
\;,
\end{split}
\end{align}
in which $J_a^+
    = f^{abc}\partial^+A^i_b A^c_i+\bar{\Psi}\gamma^+T^a\Psi$.
The instantaneous fermion vertex is given by
\begin{align}
 \begin{split}
   W_f
   =&
    \frac{g^2}{2}
   \int\diff x^-\diff^2 x_\perp
   \bar{\Psi}_c\gamma^i A^a_i T^a_{c,c'}\frac{\gamma^+}{i\partial^+}\gamma^j A^b_j T^b_{c', c''}\Psi_{c''}
\;.
 \end{split}
\end{align}

Since we are interested in the interactions introduced by the background field to the quantum state, but not the dynamics of the background field itself, we neglect the kinetic energy of the background field and its self-interaction.
Then \(V_{\mathcal{A}}(x^+)\) contains all the terms in \(V_{qg} + V_{ggg} + V_{gggg} + W_{g} + W_{f}\) that involve the gauge field by replacing one of the \(A^\mu\)s with \(\mathcal{A}^\mu\).
In the problem of interest, the only nonzero component is \(\mathcal{A}^-_a\); see more details of the background field in Ref.~\cref{sec:background_A}.
Note that in the presence of the transverse components \(\mathcal{A}^i_a\), additional background field interactions will contribute to the instantaneous terms.
The background field interaction term \(V_{\mathcal{A}}(x^+)\) therefore consists of two terms:
\begin{align}
    V(x^+)=V_{q\mathcal{A}}(x^+) + V_{g\mathcal{A}}(x^+)\;.
\end{align}
The $V_{q\mathcal{A}}$ term is the interaction between the fermion and the background gluon field, 
\begin{align}
  V_{q\mathcal{A}}=\int \diff x^- \diff^2 x_\perp
   g\bar{\Psi}\gamma^+ T^a\Psi\mathcal{A}^a_+
   \;,
\end{align}
and $V_{g\mathcal{A}}$ is the interaction between the dynamical gluon and the background gluon field, 
\begin{align}\label{eq:VgA}
  V_{g\mathcal{A}}=\int \diff x^- \diff^2 x_\perp
   g f^{abc} \partial^+ A^i_b A^c_i \mathcal{A}^a_+
   \;.
\end{align}
In the following, we present the second-quantized operators relevant under the approximation introduced in \cref{eq:jet_states}.
The resulting expressions are collected in \cref{tab:second_quantized} and \cref{tab:second_quantized_cont}, which are directly applicable to the quantum simulations of both quark and gluon jets. 
While a version of these expressions has already been provided in Chapter 4 of Ref.~\cite{Brodsky:1997de}, our aim here is to provide a concise reference consistent with the conventions chosen for the current and future studies, as well as to correct any typographical errors in the original text. We do not include the self-induced inertia operators, as their contribution can be absorbed into the mass counterterm when implementing mass renormalization \cite{Brodsky:1997de}.

The second-quantized Hamiltonian can be obtained by inserting the mode expansion of the field operators, as in \cref{eq:box_modes}, into the corresponding operators:
\begin{itemize}
  \item $P^-_\textrm{KE}$\\
  The kinetic energy terms contains that of the quark, antiquark, and gluon, respectively, 
  \begin{align}
\begin{split}
    P^-_\textrm{KE}=&K_q + K_{\bar q} + K_g
    \;,
\end{split}
\end{align}
 in which
 \begin{subequations}
 \begin{align}
    K_q= &\sum_{\ind} \frac{\vec{p}_{\perp}^2+m_q^2}{p^+} b_\ind^\dagger b^\pd_\ind\;,\\
    K_{\bar q}=& \sum_{\ind} \frac{\vec{p}_{\perp}^2+m_{ q}^2}{p^+}d_\ind^\dagger d^\pd_\ind\;,\\
    K_g=& \sum_{\ind} \frac{\vec{p}_{\perp}^2}{p^+}a_\ind^\dagger a^\pd_\ind
    \;.
\end{align}
 \end{subequations}

\item $V_{q\mathcal{A}}$

The interaction operator of the background field with the quark reads
\begin{align}
  \begin{split}
    V_{q\mathcal{A}} (x^+)
    =& g \int \diff x^-\diff^2 x_\perp
    \sum_{\ind_1,\ind_2} 
    \frac{1}{\sqrt{p_1^+ p_2^+} \Omega}
     \\
    &b^\dagger_{\ind_2}\bar{u}(p_2,\lambda_2) e^{ip_2 \cdot x}\gamma^+
    b^\pd_{\ind_1} u(p_1,\lambda_1)\\
    &e^{-ip_1\cdot x} T_{c_2,c_1}^a\mathcal{A}^a_+(\vec x_\perp , x^+)
    \;.
  \end{split}
\end{align}
The integration over $x^-$ yields the conservation of $p^+$.
The integration over $x_\perp$ performs a Fourier transformation on the background field, $\int\diff^2 x_\perp \mathcal{A}^a_+(\vec x_\perp, x^+) e^{-i\vec k_\perp\cdot \vec x_\perp} $  $=\tilde{\mathcal{A}}^a_+(\vec k_\perp, x^+)$, thereby
\begin{align}
\begin{split}
  V_{q\mathcal{A}} (x^+)
    =&\frac{2g}{\Omega_\perp }\sum_{\ind_1,\ind_2} 
    \delta_{k_2^+,k_1^+}
    \delta_{\lambda_1,\lambda_2}
     T_{c_2,c_1}^a\\
     &\tilde{\mathcal{A}}^a_+(\vec p_{2,\perp}- \vec p_{1,\perp}, x^+)
     b^\dagger_{\ind_2} 
    b^\pd_{\ind_1} 
     \;.
\end{split}
\end{align}

\item $V_{g\mathcal{A}}$ \\
The interaction operator of the background field with the gluon reads
\begin{align}
  \begin{split}
  V_{g\mathcal{A}} (x^+)
   =& 
   g\int \diff x^- \diff^2 x_\perp
   \sum_{\ind_1,\ind_2} \frac{1}{\sqrt{p_1^+ p_2^+ }\Omega}f^{a a_1 a_2} 
   \\
  & 
  \partial^+ [a^\pd_{\ind_1}\epsilon^i(p_1,\lambda_1)e^{-ip_1\cdot x}
  \\
  &
  + a^\dagger_{\ind_1}\epsilon^{* i}(p_1,\lambda_1)e^{ip_1\cdot x}] \\
  &
  \times
[a^\pd_{\ind_2}\epsilon_i(p_2,\lambda_2)e^{-ip_2\cdot x}
  \\
  &
  + a^\dagger_{\ind_2}\epsilon_i^*(p_2,\lambda_2)e^{ip_2\cdot x}]
 \mathcal{A}^a_+ (\vec x_\perp, x^+)
   \;.
  \end{split}
\end{align}
It contains four terms, $a^\dagger a$, $a a^\dagger$, $a^\dagger a^\dagger$, and $a a$. The latter two terms correspond to gluons radiated/absorbed by the background field. In the problem background field, $p^+=0$, therefore these two terms vanish by $p^+$ conservation, as the $x^-$ integration yields $\delta(p_1^+ + p_2^+)$. 
The remaining two terms combines into 
\begin{align}
  \begin{split}
  V_{g\mathcal{A}} (x^+)
=& -\frac{ i 2 g }{\Omega_\perp}\sum_{\ind_1,\ind_2} 
\delta_{k_1^+,k_2^+}
\delta_{\lambda_1,\lambda_2}\\
  &
f^{a a_2 a_1}
\tilde{\mathcal{A}}^a_+(\vec p_{2,\perp}- \vec p_{1,\perp}, x^+)
a^\dagger_{\ind_2}
a^\pd_{\ind_1}
   \;.
  \end{split}
\end{align}

\item $V_{qg}$\\
The fermion-gluon vertex interaction reads
\begin{align}
  \begin{split}
    V_{q g}
    =&   \sum_{\ind_1,\ind_2} \sum_{\ind_3} 
    \frac{g}{\sqrt{ p_1^+ p_2^+ p_3^+ \Omega^3}}
    \int \diff x^-\diff^2 x_\perp
    \\
  &b^\dagger_{\ind_2} \bar{u}(p_2,\lambda_2) e^{ip_2 \cdot x}
    \gamma^\mu T_{c_2,c_1}^{a_3}\\
    &\times
    b^\pd_{\ind_1} u(p_1,\lambda_1)e^{-ip_1\cdot x}
  [a^\pd_{\ind_3}\epsilon_\mu(p_3,\lambda_3)e^{-ip_3\cdot x}
  \\
  &+ a^\dagger_{\ind_3}\epsilon_\mu^*(p_3,\lambda_3)e^{ip_3\cdot x}]
    \;.
  \end{split}
\end{align}
We keep the two terms that contribute to the quark Fock sectors $\ket{q g^N}$, and obtain
\begin{align}\label{eq:Vqg_res}
  \begin{split}
    V_{q g}
  =&  \sum_{\ind_1,\ind_2} \sum_{\ind_3} 
    \frac{g }{\sqrt{ p_1^+ p_2^+ p_3^+ \Omega}}
   T_{c_2,c_1}^{a_3}
  \delta^{(3)}_{p_2 - p_1 +p_3}
   \\
  &\Delta_1^{2,3}
     a^\dagger_{\ind_3} b^\dagger_{\ind_2} b^\pd_{\ind_1} 
     +\text{h.c.}
    \;,
  \end{split}
\end{align}
in which
\begin{align}\label{eq:Dm}
    \Delta_1^{2,3}\equiv \bar{u}(p_2,\lambda_2)
    \slashed{\epsilon}^*(p_3,\lambda_3)
    u(p_1,\lambda_1) \;.
\end{align}
The explicit expression of $\Delta_1^{2,3}$ is listed in~\cref{tab:ggg_Sigma}.
This term corresponds to $V_1 $ in Ref.~\cite{Brodsky:1997de}.

  \item $V_{ggg}$\\
  The three-gluon vertex reads
  \begin{align}
  \begin{split}
    V_{ggg}
    =&
     g \int\diff x^-\diff^2 x_\perp
     \sum_{\ind_1, \ind_2, \ind_3} 
    f^{a_1 a_2 a_3}\\
  &
    \frac{1}{\sqrt{p_1^+ p_2^+ p_3^+\Omega^3}}
 \partial^\mu  [
    a^\pd_{\ind_1}\epsilon^i(p_1,\lambda_1)e^{-ip_1\cdot x}
  \\
  &+ a^\dagger_{\ind_1}\epsilon^{* i}(p_1,\lambda_1)e^{ip_1\cdot x}]
  \\
  &\times
   [a^\pd_{\ind_2}\epsilon_i(p_2,\lambda_2)e^{-ip_2\cdot x}
  \\
  &+ a^\dagger_{\ind_2}\epsilon_i^*(p_2,\lambda_2)e^{ip_2\cdot x}]\\
  &\times
[a^\pd_{\ind_3}\epsilon_\mu(p_3,\lambda_3)e^{-ip_3\cdot x}
  \\
  &+ a^\dagger_{\ind_3}\epsilon_\mu^*(p_3,\lambda_3)e^{ip_3\cdot x}]
 \;.
  \end{split}
 \end{align}
Here and in the following, the contributions from the pure creation and annihilation operators are excluded, as they vanish by $p^+$ conservation when $p^+=0$ modes are neglected. The remaining terms can be written collectively as
\begin{align}\label{eq:V_ggg_res}
  \begin{split}
      V_{ggg}  =&
    -ig  
    \sum_{\ind_1, \ind_2, \ind_3} 
    \frac{1}{\sqrt{p_1^+p_2^+ p_3^+ \Omega}}
    \delta^3_{p_1-p_2-p_3}\\
    &
    f^{a_1 a_2 a_3}
    \Sigma_1^{2,3}
    a^\dagger_{\ind_2}
    a^\dagger_{\ind_3}
    a^\pd_{\ind_1}   +\text{h.c.} \;,
  \end{split}
\end{align}
where we have defined a helicity structure,
    \begin{align}\label{eq:Sigma_ggg}
      \begin{split}
        \Sigma_1^{2,3}
   \equiv  
  &  
    \left[ \epsilon(p_1,\lambda_1)
     \cdot \epsilon^*(p_2,\lambda_2)\right]
     \left[ p_1\cdot \epsilon^*(p_3,\lambda_3)\right. \\
     &\left. + p_2\cdot \epsilon^*(p_3,\lambda_3)\right]\\
     &+
     \left[ \epsilon^*(p_2,\lambda_2)
     \cdot \epsilon^*(p_3,\lambda_3)\right]
     \left[ p_3\cdot \epsilon(p_1,\lambda_1)\right]
  \;.
  \end{split}
  \end{align}
The explicit expression of $\Sigma_1^{2,3}$ is listed in~\cref{tab:ggg_Sigma}.
This term corresponds to $V_{3,1-3} $ in Ref.~\cite{Brodsky:1997de}.

Analogous to the $q\to qg$ splitting process, it is convenient to define the longitudinal momentum fraction of one gluon in relative to the total gluons as $z\equiv z_3= p_3^+/p_1^+$, as such $p^+_3 = z p_1^+ $ and $p^+_2 = (1-z) p_1^+= z_2 p_1^+$. 
Define the relative momenta between the two final(initial) gluons and the initial(final) gluon as,
\begin{align}
\begin{split}
  &\vec\Delta_2 \equiv \vec p_{2,\perp}-  \vec p_{1,\perp}
  \;,
\\
&\vec\Delta_3 \equiv \vec p_{3,\perp} - ( \vec p_{2,\perp} +  \vec p_{3,\perp})
\;,
  \end{split}
\end{align}
 and
 \begin{align}
  \vec \Delta_m
  \equiv -(1-z)\vec\Delta_2
  +z \vec\Delta_3
\;.
\end{align}
The treatment of momentum conservation on the periodic lattice is in consistent with the that for the $V_{qg}$ term \cite{Li:2021zaw}.

\begin{table}[t]
\centering
  \caption{Explicit expressions of helicity structures, $\Delta_1^{2,3} $ as defined in~\cref{eq:Dm}, and $ \Sigma_1^{2,3}$ as defined in~\cref{eq:Sigma_ggg} for different helicity configurations. 
  Here, we define $z\equiv p_3^+/p_1^+$ and $ \vec \Delta_{m}
  \equiv (1-z)\vec p_{3,\perp}
  -z \vec p_{2,\perp} $.
  We inherit the treatment of momentum conservation and the evaluation of $\vec \Delta_m$ on the periodic lattice developed in Ref.~\cite{Li:2021zaw}.
  }\label{tab:ggg_Sigma}
  \begin{tabular}{c c c }
    \hline\hline
  \begin{tabular}{c} 
  Helicity configurations  \\ 
    ($\lambda_1, \lambda_2, \lambda_3$)
    \end{tabular}
    & $\Delta_1^{2,3} $
    & ~~~~~
   $ \Sigma_1^{2,3}$~~~~~~~~~~
    \\
    \hline
    $\uparrow\uparrow\uparrow$   
      & $ \dfrac{\sqrt{2}}{z\sqrt{1-z}}\Delta_m^L$
&  $-\dfrac{\sqrt{2}}{z}\Delta_m^L$
\\
$\uparrow\uparrow\downarrow$    
& $ \dfrac{\sqrt{2(1-z)}}{z} \Delta_m^R$
&  $\dfrac{-2+z}{\sqrt{2}z}\Delta_m^R$
\\
$\uparrow\downarrow\uparrow$    
& $ -\dfrac{z\sqrt{2}}{\sqrt{1-z}} m_q$
&
$\dfrac{1}{\sqrt{2}}\Delta_m^R$
\\
$\uparrow\downarrow\downarrow$    
& 0
&0
\\
$\downarrow\uparrow\uparrow$  
& 0
&0
\\
$\downarrow\uparrow\downarrow$    
& $\dfrac{z\sqrt{2}}{\sqrt{1-z}} m_q$
&$\dfrac{1}{\sqrt{2}}\Delta_m^L$
\\
$\downarrow\downarrow\uparrow$    
& $ \dfrac{\sqrt{2(1-z)}}{z} \Delta_m^L$
 &  $\dfrac{-2+z}{\sqrt{2}z}\Delta_m^L$
\\
$\downarrow\downarrow\downarrow$  
& $ \dfrac{\sqrt{2}}{z\sqrt{1-z}} \Delta_m^R$

&  $-\dfrac{\sqrt{2}}{z}\Delta_m^R$
\\
\hline\hline
\end{tabular}
\end{table}

  \item $V_{gggg}$\\
  The four-point-gluon vertex reads,
  \begin{align}
    \begin{split}
      V_{gggg}
      =&\frac{g^2}{4}  \int\diff x^-\diff^2 x_\perp 
      \sum_{\ind_1, \ind_2, \ind_3, \ind_4} \\
      &
      f^{a a_1 a_2}  f^{a a_3 a_4}
      \frac{1}{\sqrt{ p_1^+ p_2^+p_3^+ p_4^+\Omega^4}}\\
      &
      [a^\pd_{\ind_1}\epsilon^i(p_1,\lambda_1)e^{-ip_1\cdot x}+a^\dagger_{\ind_1}\epsilon^{i *}(p_1,\lambda_1)e^{ip_1\cdot x}]\\
      & 
      [a^\pd_{\ind_2}\epsilon^j(p_2,\lambda_2)e^{-ip_2\cdot x}+a^\dagger_{\ind_2}\epsilon^{j *}(p_2,\lambda_2)e^{ip_2\cdot x}] \\
      &
      [a^\pd_{\ind_3}\epsilon_i(p_3,\lambda_3)e^{-ip_3\cdot x}+a^\dagger_{\ind_3}\epsilon_i^{ *}(p_3,\lambda_3)e^{ip_3\cdot x}] \\
      &
      [a^\pd_{\ind_4}\epsilon_j(p_4,\lambda_4)e^{-ip_4\cdot x}+a^\dagger_{\ind_4}\epsilon_j^{ *}(p_4,\lambda_4)e^{ip_4\cdot x}] 
   \;.
    \end{split}
   \end{align}
The containing terms can be collected into the following two parts.
The 2-to-2 interaction is written as 
\begin{align}
  \begin{split}
    V_{gggg,1}
    =& \frac{g^2}{2} 
    \sum_{\ind_1, \ind_2, \ind_3, \ind_4} 
    \frac{\delta^3_{p_1+p_2-p_3-p_4}}{\sqrt{ p_1^+ p_2^+p_3^+ p_4^+} \Omega}
    \bigg[\\
      &
    f^{a a_1 a_2}  f^{a a_3 a_4} \delta_{\lambda_1, \lambda_3}
    \delta_{\lambda_2, \lambda_4}
    \\
    +&f^{a a_1 a_3}  f^{a a_2 a_4}  \delta_{\lambda_1, -\lambda_2}
    \delta_{\lambda_3, -\lambda_4}
    \\
      &+f^{a a_1 a_4}  f^{a a_3 a_2} \delta_{\lambda_1, \lambda_3}
    \delta_{\lambda_2, \lambda_4}
    \bigg]\\
      &a^\dagger_{\ind_4} a^\dagger_{\ind_3} 
     a^\pd_{\ind_2} 
         a^\pd_{\ind_1} 
 \;,
  \end{split}
 \end{align}
which corresponds to $F_{9,3-5} $ in Ref.~\cite{Brodsky:1997de},

 The 1-to-3 and 3-to-1 interaction is written as 
 \begin{align}
   \begin{split}
     V_{gggg,2}
=& g^2
\sum_{\ind_1, \ind_2, \ind_3, \ind_4} 
\frac{1}{\sqrt{ p_1^+ p_2^+p_3^+ p_4^+} \Omega}\\
      &
\delta^3_{p_1-p_2-p_3-p_4}
f^{a a_1 a_2}  f^{a a_3 a_4} \delta_{\lambda_1, \lambda_3}
\delta_{\lambda_2, -\lambda_4}\\
      &
a^\dagger_{\ind_4} a^\dagger_{\ind_3} 
     a^\dagger_{\ind_2} 
         a^\pd_{\ind_1} 
+\text{h.c.}
  \;,
   \end{split}
  \end{align}
which corresponds to $F_{9,2} $ in Ref.~\cite{Brodsky:1997de}.

\item $ W_{g,g}$\\
The pure gluon part of the instantaneous gluon vertex can be written as
\begin{align}
  \begin{split}
    W_{g,g}=&\frac{g^2}{2}
     \int\diff x^-\diff^2 x_\perp
     f^{abc}\partial^+A^i_b A^c_i \frac{1}{{(i\partial^+)}^2}\\
     &
     f^{aef}\partial^+A^j_e A^f_j
  \;.
  \end{split}
  \end{align}
The containing terms can be collected into the following three pieces,
       \begin{align}
        \begin{split}
          W_{g,1}
       =&- g^2
       \sum_{\ind_1, \ind_2, \ind_3, \ind_4} 
       \delta^3_{p_1+p_2+p_3-p_4}\\
       &
       \frac{1}{\sqrt{ p_1^+ p_2^+p_3^+ p_4^+}\Omega}
       f^{a a_1 a_2} f^{a a_3 a_4}\\
       &
        \frac{ p_1^+ (p_3^+ + p_4^+)}{{(p_3^+ -p_4^+ )}^2}
       \delta_{\lambda_1, -\lambda_2}
       \delta_{\lambda_3, \lambda_4}\\
       &
       a_{\ind_4}^\dagger a^\pd_{\ind_1} a^\pd_{\ind_2} a^\pd_{\ind_3} 
       +\text{h.c.}
    \;,
    \end{split}
    \end{align}
    which corresponds to $F_{9,1} $ in Ref.~\cite{Brodsky:1997de},
        \begin{align}
      \begin{split}
        W_{g,2}
        =&g^2
        \sum_{\ind_1, \ind_2, \ind_3, \ind_4} 
        \delta^3_{p_1+p_2-p_3-p_4}
        \frac{1}{\Omega}\\
       &
        \sqrt{ \frac{p_1^+ p_3^+ }{ p_2^+ p_4^+} }
        f^{a a_1 a_2} f^{a a_3 a_4}
     \frac{1}{{(p_3^+ +p_4^+ )}^2}\\
       &
     \delta_{\lambda_1, -\lambda_2}
     \delta_{\lambda_3, -\lambda_4}
    a^\dagger_{\ind_3}  a^\dagger_{\ind_4} a^\pd_{\ind_1} a^\pd_{\ind_2} 
      \;,
      \end{split}
      \end{align}
   which corresponds to $S_{9,2} $ in Ref.~\cite{Brodsky:1997de}, and 
      \begin{align}
        \begin{split}
          W_{g,3}
          =& -\frac{g^2}{2}
           \sum_{\ind_1, \ind_2, \ind_3, \ind_4} 
           \delta^3_{p_1+p_2-p_3-p_4}\\
       &
           \frac{1}{\sqrt{ p_1^+ p_2^+p_3^+ p_4^+} \Omega}
           f^{a a_1 a_3} f^{a a_2 a_4}
           \\
           &
          \frac{(p_1^+ + p_3^+)( p_2^+ + p_4^+)}{{(p_2^+ -p_4^+ )}^2}
          \delta_{\lambda_1, \lambda_3}
          \delta_{\lambda_2, \lambda_4}\\
       &
          a_3^\dagger a_4^\dagger  a_1  a_2 
        \;,
        \end{split}
        \end{align}
   which corresponds to $S_{9,1} $ in Ref.~\cite{Brodsky:1997de}.

   \item $ W_{g,q}$\\
The quark-gluon part of the instantaneous gluon vertex can be written as
    \begin{align}
      \begin{split}
        W_{g,qg}
     =&\frac{g^2}{2}
     \int\diff x^-\diff^2 x_\perp
     \bigg[\\
     &
     f^{abc}\partial^+A^i_b A^c_i \frac{1}{{(i\partial^+)}^2}
     \bar{\Psi}_q\gamma^+T^a\Psi_q\\
     &
     +  \bar{\Psi}_q\gamma^+T^a\Psi_q\frac{1}{{(i\partial^+)}^2}f^{abc}\partial^+A^i_b A^c_i 
     \bigg]
  \;,
  \end{split}
  \end{align}
  in which the field operator $ \Psi_q$ ($\bar{\Psi}_q$) contains only the  $b^\dagger $ ( $b$ ) part of $ \Psi$ ($\bar{\Psi}$) provided in \cref{eq:box_modes}.

  We collect the containing terms into,
\begin{align}
  \begin{split}
    W_{g,4}
=&2 i g^2
\sum_{\ind_1, \ind_2, \ind_3, \ind_4} 
\delta^3_{p_1+p_2+p_3-p_4}
\frac{1}{\Omega}\\
    &
\sqrt{ \frac{p_1^+} {p_2^+}}
f^{a a_1 a_2} T^a_{c_4, c_3}
\frac{1}{{(p_3^+ -p_4^+)}^2}\\
&
\delta_{\lambda_1,-\lambda_2}
\delta_{\lambda_3,\lambda_4}
b^\dagger_{\ind_4}  b^\pd_{\ind_3}  a^\pd_{\ind_1} a^\pd_{\ind_2}
+\text{h.c.}
\;,
\end{split}
\end{align}
   which corresponds to $F_{5,2} $ in Ref.~\cite{Brodsky:1997de}, and 
   \begin{align}
  \begin{split}
    W_{g,5}
    =&2 i g^2
    \sum_{\ind_1, \ind_2, \ind_3, \ind_4} 
    \delta^3_{p_1+p_2-p_3-p_4}\\
    &
    \frac{p_1^+ +p_3^+ }{{(p_4^+ - p_2^+)}^2}
    \frac{1}{\sqrt{ p_1^+ p_3^+ }\Omega}
    f^{a a_1 a_3} T^a_{c_4, c_2}\\
    &
   \delta_{\lambda_1,\lambda_3}
   \delta_{\lambda_2,\lambda_3}
   a^\dagger_{\ind_3} b^\dagger_{\ind_4}   a^\pd_{\ind_1} b^\pd_{\ind_2} 
\;,
\end{split}
\end{align}
 which corresponds to $F_{5,3} $ in Ref.~\cite{Brodsky:1997de}.

    \item $ W_{f,q}$\\
The quark part of the instantaneous fermion vertex can be written as
\begin{align}
  \begin{split}
    W_{f,q}
    =&
        \frac{g^2}{2}
    \int\diff x^-\diff^2 x_\perp
    \bar{\Psi}_{q,c_1}\gamma^i A^a_i T^a_{c_1,c}\\
    &\frac{\gamma^+}{i\partial^+}\gamma^j A^b_j T^b_{c, c_4}\Psi_{q,c_4}
 \;.
  \end{split}
 \end{align}
We write it into three terms,
  \begin{align}
    \begin{split}
      W_{q,1}
       =&
       2 g^2
      \sum_{\ind_1, \ind_2, \ind_3, \ind_4} 
      \delta^3_{p_2+p_3+p_4-p_1}\\
      &
      \frac{1}{\sqrt{ p_3^+ p _4^+}\Omega}
      T^{a_3}_{c_2,c}  T^{a_4}_{c, c_1}
      \frac{1}{p_1^+ - p_4^+}\\
      &
      \delta_{\lambda_2,\lambda_3}
      \delta_{\lambda_4,-\lambda_1}
      \delta_{\lambda_2,\lambda_1}
       b^\dagger_{\ind_2}  a^\dagger_{\ind_3}  a^\dagger_{\ind_4} b^\pd_{\ind_1}
       +\text{h.c.}
   \;,
    \end{split}
   \end{align}
 which corresponds to $F_{5,1} $ in Ref.~\cite{Brodsky:1997de},
        \begin{align}
       \begin{split}
         W_{q,2}
          =&
          2g^2
          \sum_{\ind_1, \ind_2, \ind_3, \ind_4} 
          \delta^3_{p_3-p_2+p_4-p_1}\\
      &
          \frac{1}{\sqrt{  p_2^+ p_4^+ }\Omega}
          T^{a_2}_{c_3,c}  T^{a_4}_{c, c_1}
          \frac{1}{p_1^+ - p_4^+}\\
      &
          \delta_{\lambda_3,-\lambda_2}
         \delta_{\lambda_3,-\lambda_4}
         \delta_{\lambda_3,\lambda_1}
           b^\dagger_{\ind_3} a^\dagger_{\ind_4} a^\pd_{\ind_2}  b^\pd_{\ind_1} 
      \;.
       \end{split}
      \end{align}
       which corresponds to $S_{5,1} $ in Ref.~\cite{Brodsky:1997de}, and 
     \begin{align}
       \begin{split}
         W_{q,3}
          =&
          2g^2
         \sum_{\ind_1, \ind_2, \ind_3, \ind_4} 
         \delta^3_{p_3+p_4-p_2-p_1}\\
      &
         \frac{1}{\sqrt{  p_4^+ p_2^+ }\Omega}
        T^{a_4}_{c_3,c}  T^{a_2}_{c, c_1}
        \frac{1}{p_2^+ + p_1^+}\\
      &
        \delta_{\lambda_3,\lambda_4}
       \delta_{\lambda_3,\lambda_2}
       \delta_{\lambda_3,\lambda_1}
         b^\dagger_{\ind_3} a^\dagger_{\ind_4} a^\pd_{\ind_2} b^\pd_{\ind_1} 
      \;,
       \end{split}
      \end{align}
      which corresponds to $S_{5,2} $ in Ref.~\cite{Brodsky:1997de}.
\end{itemize}

\bibliographystyle{apsrev4-1}
\bibliography{main.bib}

\end{document}